\documentclass[aps,prd,reprint, superscriptaddress, twocolumn,nofootinbib]{revtex4-2}

\usepackage{amsmath,amssymb,amsthm,amstext}
\usepackage{natbib}
\usepackage{graphicx}
\usepackage{color}
\usepackage{array, enumerate}
\usepackage{bm}
\usepackage{multirow}
\usepackage[breaklinks,colorlinks,citecolor=blue]{hyperref}
\usepackage{braket}
\usepackage{txfonts}

\def\nn{\nonumber}
\def\be{\begin{equation}}
\def\ee{\end{equation}}

\definecolor{Green}{rgb}{0.0, 0.5, 0.0}

\begin{document}
\title{Resonant dynamics of extreme mass-ratio inspirals in a perturbed Kerr spacetime}
\author{Zhen Pan}
\email{zhpan@sjtu.edu.cn}
\affiliation{Tsung-Dao Lee Institute, Shanghai Jiao-Tong University, Shanghai, 520 Shengrong Road, 201210, People’s Republic of China}
\affiliation{School of Physics $\&$ Astronomy, Shanghai Jiao-Tong University, Shanghai, 800 Dongchuan Road, 200240, People’s Republic of China}
\affiliation{Perimeter Institute for Theoretical Physics, Ontario, N2L 2Y5, Canada}
\author{Huan Yang}
\email{hyang@perimeterinstitute.ca}
\affiliation{Perimeter Institute for Theoretical Physics, Ontario, N2L 2Y5, Canada}
\affiliation{University of Guelph, Guelph, Ontario N1G 2W1, Canada}
\author{Laura Bernard}
\affiliation{Laboratoire Univers et Théories, Observatoire de Paris, Université PSL, Université Paris Cité, CNRS, F-92190 Meudon, France}
\author{B\'eatrice Bonga}
\affiliation{Institute for Mathematics, Astrophysics and Particle Physics, Radboud University, 6525 AJ Nijmegen, The Netherlands}

\begin{abstract}
Extreme mass-ratio inspirals (EMRI) are one of the most sensitive probes of black hole spacetimes with gravitational wave measurements. In this work, we systematically analyze the dynamics of an EMRI system near orbital resonances, assuming the background spacetime is weakly perturbed from Kerr. Using the action-angle formalism, we have derived an effective resonant Hamiltonian that describes the dynamics of the resonant degree of freedom, for the case that the EMRI motion across  the  resonance regime.
This effective resonant Hamiltonian can also be used to derive the condition that the trajectory enters/exits a resonant island and the permanent change of action variables across the resonance with the gravitational wave radiation turned on. The orbital chaos, on the other hand, generally leads to transitions between different branches of rotational orbits with finite changes of the action variables. These findings are demonstrated with numerical orbital evolutions that are mapped into representations using action-angle variables. This study is one  part of the program of understanding EMRI dynamics in a generic perturbed Kerr spacetime, which paves the way of using EMRIs to precisely measure the black hole spacetime. 
\end{abstract}

\maketitle

\section{Introduction}
An extreme mass-ratio inspiral (EMRI) system comprises a supermassive black hole and a stellar-mass compact object, i.e., a black hole or a neutron star \cite{Babak:2017tow}. Together with massive black hole binaries, EMRIs  are commonly believed to be the main extragalactic transient gravitational-wave sources for spaceborne gravitational wave detectors, such as LISA, Taiji and Tianqin \cite{Baker:2019nia,Hu:2017mde,TianQin:2020hid}.  Their formation can be classified into two distinct channels: one associated with multi-body scattering in nuclear star clusters (``dry channel") \cite{Babak:2017tow} and the other associated with accretion-assisted migration (``wet channel") \cite{Kocsis:2011dr,Yunes:2011ws,Barausse:2014tra}. It has been shown that accretion disks dramatically boots the EMRI formation rate, such that the wet EMRIs maybe more common for space-borne gravitational-wave detection \cite{Pan:2021oob,Pan:2021ksp}. There are other formation mechanisms proposed, such as the destruction of a stellar-mass black hole binary in the vicinity of a supermassive black hole \cite{ColemanMiller:2005rm,Chen:2018axp}, but the rate is rather uncertain. Recently, there is a proposal suggesting enhanced EMRI formation rates in near supermassive black hole binaries \cite{Naoz:2022rru}, although concerns have been raised regarding the supply of stellar-mass black holes \cite{Mazzolari:2022cho}.

EMRIs have a wide range of astrophysical applications. First, as the mass and spin of the host massive black hole can be measured accurately (down to percent-level or better), a catalog of EMRI events may be used to infer the distribution of massive black holes within $10^5-10^7 {\rm M}_\odot$, which helps to understand the growth mechanisms of massive black holes. Secondly, as wet EMRIs are generally accompanied by active galactic nuclei (AGN), they are ideal candidates for multi-messenger observations, which are particularly useful for studying the accretion physics. Thirdly, the less-massive object in an EMRI system may be a mass-gap object \cite{Pan:2021lyw,Pan:2021xhv} (similar to the one detected in GW190814) or a primordial black hole \cite{Barsanti:2021ydd}, so that the EMRI observations may be used to probe the existence of these objects, determining their abundance and diagnosing their formation mechanisms.

EMRIs also have important applications in fundamental physics, including testing strong-field predictions of General Relativity. They generally have a superior power in detecting weak environmental forces because of the large number ($10^4-10^5$) of orbital cycles in-band, so that weak effects may be amplified to achieve detectable gravitational wave phase shifts. These environmental forces may come from the tidal gravitational field of a third stellar-mass object \cite{Yang:2017aht,Bonga:2019ycj,Gupta:2021cno,Gupta:2022fbe,Camilloni:2023rra},
the migration force from an accretion disk and/or the interaction between  the stellar-mass black hole and a possible  dark-matter cloud \cite{Ferreira:2017pth,Hannuksela:2018izj,Baumann:2018vus,Zhang:2019eid,Zhang:2018kib,Tomaselli:2023ysb,Vicente:2022ivh,Traykova:2021dua}.
From the perspective of testing General Relativity, 
it is interesting to test the Kerr metric as a key prediction of General Relativity for rotating black holes, as modified gravity theories may predict different black hole spacetimes (e.g. Einstein-dilaton-Gauss-Bonnet \cite{Ayzenberg:2014aka}, Dynamical Chern-Simons Gravity \cite{yunes2009dynamical} and effective field theory extensions of General Relativity \cite{Cardoso:2018ptl}). If the central body is a black hole mimicker such as a boson star, a gravastar and/or a wormhole \cite{Cardoso:2019rvt} (despite possible issues with stability \cite{Poisson:1995sv,Dias:2010uh,Armendariz-Picon:2002gjc,Yang:2022gic}), the external metric may also be different from Kerr. 
A related, important question is: If the background spacetime is $g=g_{\rm Kerr}+h$ (assuming $|h|\ll 1$), how do we use EMRIs to probe/constrain $h$?   Notice that the presence of $h$ not only modifies the EMRI dynamics as an additional force, but also the radiated flux with respect to the same trajectory.

To understand the EMRI evolution within the background spacetime described by the metric $g=g_{\rm Kerr}+h$ (assuming $h$ is stationary), we can separate the EMRI orbit into non-resonant and resonant regimes. In the non-resonant regime, the metric perturbation $h$ introduces an extra conservative force, and the EMRI orbit oscillates around the Kerr geodesics according to the Kolmogorov–Arnold–Moser (KAM) theorem, i.e., there is no orbital chaos. With the gravitational wave radiation included, we can intuitively argue that the conservative quantities are relatively shifted by $\mathcal{O}(h)$ and the radiated flux is relatively modified by $\mathcal{O}(h)$, so that the resulting overall gravitational wave phase shift is
\begin{align}
\delta \Psi_{\rm non-res} \sim \frac{1}{q}\times \mathcal{O}(h)
\end{align}
where $q$ ($\sim 10^{-4} -10^{-6}$) is the EMRI mass ratio and $1/q$ represents the number of cycles in-band. Notice that although both $q,|h|\ll 1$, $|h|$ can be larger than $q$ so that $\delta \Psi \ge 1$, which will be observable by space-borne detectors. Here, we will be particularly interested in the scenario with $|h|>q$. In order to describe the long-term secular evolution in the non-resonant regime, one needs to work out the modified radiation flux with the modified Teukolsky equation, and determine how to relate the radiated flux to orbital quantities in the modified spacetime. We shall present this part of the analysis in a separate work.

In the resonant regime, i.e., $k \Omega^r+m \Omega^\phi+n \Omega^\theta \approx 0$ for $k,m,n \in \mathbb{Z}$ where $\Omega^{r,\theta,\phi}$ are geodesic orbital frequencies in $r,\theta,\phi$ directions, the KAM theorem no longer applies and orbital chaos can occur. In fact,  signatures of chaotic orbits near resonances have been observed in studies of various modified Kerr spacetimes \cite{Deich:2022vna,Bronicki:2022eqa,Destounis:2020kss,Destounis:2021mqv,Destounis:2021rko,Destounis:2023gpw,Lukes-Gerakopoulos:2017jub,
Apostolatos:2009vu,Lukes-Gerakopoulos:2010ipp,Cardenas-Avendano:2018ocb,Zelenka2020,Destounis:2023khj}, e.g. in the vertical jumps of rotation numbers and volume-filling features in the phase space of the trajectory. A plateau in rotation number is sometimes observed, which should be associated with the resonant islands in the phase space. Despite this progress in understanding the phenomenology of EMRI resonant behavior in various specific spacetimes, a mathematical, universal framework for generic perturbed Kerr spacetimes is still lacking. More importantly, in order to allow gravitational wave measurements to probe the spacetime perturbation $h$, 
one needs to assess {\it the impact of $h$ in the resonant regime for the long-term EMRI evolution}, which is an important goal of this work.

Since we are working in the regime that the perturbative force (due to $h$) is greater than the gravitational radiation reaction, we first solve the conservative dynamics of the associated spacetime $g_{\rm Kerr}+h$, and view the radiation reaction as a mapping between different geodesics associated with $g_{\rm Kerr}+h$.  Using a method similar to the treatment of relativistic mean motion resonance in \cite{Yang:2019iqa}, which in turn traces back to the analysis of sustained resonance for EMRIs with self-force considered (especially the procedure of applying Near Identity Transformations) \cite{vandeMeent:2013sza}, we derive a general effective resonant Hamiltonian of the form
\begin{align}\label{eq:he1}
\mathcal{H}_{\rm eff} = \alpha_0 \Theta +\beta_0 \Theta^2 + \epsilon\sum_k H_k e^{ik Q}
\end{align}
with $\Theta, Q$ being the canonical variables for the resonant degree of freedom (DOF), and relevant definition of other variables explicitly given in Sec.~\ref{sec:Heff}.   This Hamiltonian governs the essential dynamics of the EMRI system within the resonant islands (commonly referred to as the ``libration" regime for planetary systems), and it applies for generic metric perturbations $h$, including all the specific examples mentioned in previous studies \cite{Deich:2022vna,Bronicki:2022eqa,Destounis:2020kss,Destounis:2021mqv,Destounis:2021rko,Destounis:2023gpw,Lukes-Gerakopoulos:2017jub,
Apostolatos:2009vu,Lukes-Gerakopoulos:2010ipp,Cardenas-Avendano:2018ocb,Zelenka2020,Destounis:2023khj}. The physical essence of this effective Hamiltonian --- similar to the cases of mean motion resonances widely studied in planetary systems --- is that there is a single resonant DOF that is slowly varying compared to other DOFs, so that in a ``slow timescale" where other DOFs are averaged out, the system is described by such Hamiltonian. 
When the gravitational radiation reaction is turned on, the equations of motion for the resonant DOF should be correspondingly modified. Note that away from the chaotic regime, the discussion presented in \cite{vandeMeent:2013sza} becomes particularly useful for analyzing the system's dynamics.~\footnote{The original analysis and the scaling laws presented in \cite{vandeMeent:2013sza} are developed for describing the self-force effect, but the method can be easily adapted to this work.} There are in general two outcomes as a system passes through the resonance regime under the influence of radiation reaction: transient passing and resonant trapping (dubbed as ``sustained resonance" in \cite{vandeMeent:2013sza}). The transient passing produces a long-term phase shift  
\begin{align}
\delta \Psi_{\rm transient} \sim \frac{\mathcal{O}(h)}{q^{3/2}}\,,
\end{align}
which is larger than the non-resonant effects in the formal expansions. However, this does not necessarily mean that $\delta \Psi_{\rm transient}$ is numerically greater than $\delta \Psi_{\rm non-res}$, because the values of the numerical coefficients play an important role in the overall amplitude, and these coefficients correspond to different terms in the harmonics expansion which could differ by orders of magnitude (see Eq.~\eqref{eq:h}). 

In orbits near the resonant islands, the resonant angle $Q$ is no longer bounded (the ``rotation" regime), which couples with other non-resonant DOFs and gives rise to chaotic zones in the phase space. In Chapter 9.5 of \cite{murray1999solar}, a simple pendulum problem was discussed in the Hamiltonian language to illustrate the properties of chaos using ``the Standard Map" method. It is plausible that chaos arises due to similar reasons here in the EMRI system, where the non-resonant terms effectively introduce time-dependent harmonics in the Hamiltonian. The resulting chaos may be analyzed using the (standard) algebraic map method.
The width of chaotic zones is expected to be proportional to $|h|^{1/2}$, within which the orbital motion is chaotic and volume-filling in the phase space. In traditional plots for the rotation number $\nu$ (which is defined 
as the ratio of average frequencies in two different directions, e.g., $\braket{\Omega^r}/\braket{\Omega^\theta}$), 
the chaotic zones correspond to a vertical discontinuity in the rotational number, as discussed in, e.g., \cite{Cardenas-Avendano:2018ocb,Deich:2022vna}. The chaotic orbits in general contribute $\mathcal{O}(h^{1/2})$  changes in the action variables $J_\alpha$ during transitions between different branches of the orbit (Figs.~\ref{fig:eff_O2} and \ref{fig:phase_O2}), so that the long-term impact on the gravitational wave phase is 
\begin{align}
\delta \Psi_{\rm chaos} \propto \frac{\mathcal{O}(h^{1/2})}{q}\,,
\end{align} 
In principle, the chaotic orbit may also introduce transitions into the resonant islands, after which a sustained resonance is achieved. 
One criterion for generic resonance capture studied in a planetary system is that the capture only happens in the case of ``converging" evolution, i.e., the ratio between the magnitude of frequencies is converging (towards one) in time \cite{murray1999solar}. However, the EMRI evolution is ``diverging". 
Therefore, we believe that resonance capture into resonant islands is unlikely. This point of course requires further numerical confirmation in the future. 

To test the effective Hamiltonian description, we have developed numerical algorithms to do the full evolution across resonances. In particular, we have developed numerical algorithms to map the EMRI evolution using physical variables $(r,\theta,\phi, p_r,r_\theta,p_\phi)$ to action-angle variables $(q_r, q_\theta, q_\phi, J_r, J_\theta, J_\phi)$. A direct evolution using action-angle variables is thus possible, but in realistic implementations much more susceptible to computational errors because of the numerical transformations. As a result, we choose to numerically compute the long-term evolution of EMRIs near resonance using the physical variables, and map them to action-angle variables from time-to-time in order to compare to the analysis using the effective resonant Hamiltonian. The details of the transformation between the physical and action-angle variables  are explained in Appendix~\ref{sec:app2} and \ref{sec:app3}. The numerical algorithms and the effective Hamiltonian description should apply to general perturbed Kerr metrics.
In this work, we use the spinning black hole solution in 
quadratic gravity~\cite{Dong:2021yss}  as an example, where the perturbed Kerr spacetime is still stationary and axisymmetric,
and we indeed find decent agreement between the effective Hamiltonian description and the fully numerical evolution. 

The structure of this article is organized as follows. In Section~\ref{sec:Heff}, we first introduce 
action-angle variables and their usage in Hamiltonian systems, then explain why resonances reduce the number 
of DOFs, and near-resonance orbits are governed by an effective Hamiltonian.
In Section~\ref{sec:nu}, we numerically evolve a number of near-resonance ($2/3$ resonance with $\Omega^r/\Omega^\theta\approx2/3$) orbits in physical coordinates 
$(x,p)$ and illustrate various features of these orbits in terms of rotation curves and Poincar\'e maps.
We then map $(x, p)$ to action-angle variables $(q, J)$, and to resonant variables $(Q, \Theta)$ via a Near Identity Transformation (NIT), and explain the features found using the effective Hamiltonian description.
In Section~\ref{sec:impact}, we calculate the impact of crossing resonances on the EMRI waveform using the effective Hamiltonian description. We conclude this paper in Section~\ref{sec:conclusion}. 
In Appendix~\ref{sec:app1}, we analyze the $1/2$ resonance, compare it with the $2/3$ resonance
and illustrate the general features of crossing resonances.
In Appendix~\ref{sec:app2}, we summarize the necessary steps for mapping the physical coordinates to the action-angle variables. In Appendix~\ref{sec:app3}, we show the details of performing NITs.

In this work, we use the geometric units $G=c=1$, and set the supermassive black hole mass to $M=1$ if not specified otherwise.

\section{Effective Resonant Hamiltonian}\label{sec:Heff}

Let us consider a point particle's motion in the spacetime with metric $g=g_{\rm Kerr}+h$. The total Hamiltonian is given by
\begin{align}
\mathcal{H} = \frac{1}{2} g^{\alpha\beta} p_\alpha p_\beta \; ,
\end{align}
and the magnitude of the total Hamiltonian is $\mathcal{H}= -\mu^2/2$,
where $\mu$ is the particle's rest mass and the physical coordinates  $\{ x^\alpha, p_\alpha\}$ are  canonical variables of the system. In order to separate out the dynamical effect due to $h$, we shall write the Hamiltonian as
\begin{align}
\mathcal{H} & = \frac{1}{2} g_{\rm Kerr}^{\alpha\beta} p_\alpha p_\beta +\frac{\epsilon}{2} h^{\alpha\beta} p_\alpha p_\beta \nonumber \\
&:= \mathcal{H}_{\rm Kerr} +\epsilon \, \mathcal{H}_{\rm int} \; ,
\end{align}
where $h^{\alpha \beta} = -g^{\alpha \mu}_{\rm Kerr} g^{\beta \nu}_{\rm Kerr} h_{\mu\nu}$ and $\epsilon$ is a bookkeeping index. Considering the case with $h=0$, i.e., the Kerr spacetime, one can generally 
find a canonical transformation to map the physical coordinates to the action-angle variables:
\begin{align}\label{eq:canJq}
\mathcal{J}_\alpha = \mathcal{J}_\alpha(\{ x^\beta, p_\beta\}),\quad q^\alpha = q^\alpha(\{ x^\beta, p_\beta\})\,.
\end{align}
Even though this canonical transformation 
is obtained assuming the spacetime is Kerr, in the general case that $h$ is nonzero, it still gives rise to a set of canonical variables $\{\mathcal{J}_\alpha,q^\alpha\}$ for the total Hamiltonian $\mathcal{H}$.~\footnote{However, in general, $\{\mathcal{J}_\alpha,q^\alpha\}$ are not \emph{action-angle} variables for the full Hamiltonian. } In other words, $\{\mathcal{J}_\alpha,q^\alpha\}$ satisfy
\begin{align}
\frac{d q^\alpha}{d \tau} & = \frac{\partial \mathcal{H}}{\partial \mathcal{J}_\alpha} = \frac{\partial \mathcal{H}_{\rm Kerr}}{\partial \mathcal{J}_\alpha} +\epsilon\frac{\partial \mathcal{H}_{\rm int}}{\partial \mathcal{J}_\alpha} , \nonumber \\  
\frac{d \mathcal{J}_\alpha}{d \tau} & = -\frac{\partial \mathcal{H}}{\partial q^\alpha} =-\frac{\partial \mathcal{H}_{\rm Kerr}}{\partial q^\alpha}-\epsilon \frac{\partial \mathcal{H}_{\rm int}}{\partial q^\alpha}\,,
\end{align}
where $\tau$ is the proper time of the particle. In addition, with the inverse transformation of Eq.~\eqref{eq:canJq}, $\mathcal{H}_{\rm Kerr}(\{ x^\alpha, p_\alpha\})$
may be re-written as $\mathcal{H}_{\rm Kerr}(\{\mathcal{J}_\alpha,q^\alpha\})$. Based on the definition of action-angle variables, the above equations can then be simplified as
\begin{align}\label{eq:eom}
\frac{d q^\alpha}{d \tau} &  = \Omega^\alpha +\epsilon\frac{\partial \mathcal{H}_{\rm int}}{\partial \mathcal{J}_\alpha} , \nonumber \\  
\frac{d \mathcal{J}_\alpha}{d \tau} &  =-\epsilon \frac{\partial \mathcal{H}_{\rm int}}{\partial q^\alpha}\,,
\end{align}
where $\Omega^\alpha (\{\mathcal{J}_\beta \}) :=\frac{\partial \mathcal{H}_{\rm Kerr}}{\partial \mathcal{J}_\alpha}$ are the ``angular frequencies"  for each angle variable in the Kerr spacetime.

\subsection{Effective Resonant Hamiltonian}\label{sec:subeff}

It is straightforward to see that $h$ drives the evolution of $\mathcal{J}_\alpha$, according to Eq.~\eqref{eq:eom}.
As $\mathcal{H}_{\rm int}$ is a function of $\{\mathcal{J}_\alpha,q^\alpha\}$, we Fourier decompose it as
\begin{align}\label{eq:h}
\mathcal{H}_{\rm int} = \sum_{k,m,n} H_{k,m,n}(\{\mathcal{J}_\alpha\}) e^{i(k q^r+m q^\phi+n q^\theta)}
\end{align}
where $k,m,n \in \mathbb{Z}$ and $H_{k,m,n}$ is a function of $\{\mathcal{J}_\alpha\}$. As $\dot{q}^\alpha \sim \Omega^\alpha +\mathcal{O}(\epsilon)$, the variation of $\mathcal{J}_\alpha$ is driven by oscillatory forcing terms with different frequencies. In the long term, these oscillatory forcing terms introduce no secular effects so that the trajectory varies around Kerr geodesics, and there is no chaos  in the non-resonant regime (as to be expected from the KAM theorem).
This observation also justifies using $\mathcal{J}_\alpha$ for the long-term evolution of EMRIs with gravitational radiation reaction turned on: in the non-resonant regime $h$ only generates oscillations of $\mathcal{J}_\alpha$ around its mean value $\langle \mathcal{J}_\alpha \rangle$, whereas the radiation reaction produces secular change of $\mathcal{J}_\alpha$.

In the resonant regime where the orbital frequencies are commensurate $N_\alpha \Omega^\alpha \approx 0$ for certain ${\bf N}=(N_r, N_\theta, N_\phi)$, we rewrite the equations of motion for $\{\mathcal{J}_\alpha,q^\alpha\}$ as
\begin{align}\label{eq:niteom}
\frac{d q^\alpha}{d \tau} &  = \Omega^\alpha +\epsilon \sum_{k \in \mathbb{Z}} \frac{\partial H_{k N_j}}{\partial \mathcal{J}_\alpha} e^{i k N_j q^j}+\epsilon \sum_{n_j \in R} \frac{\partial H_{n_j}}{\partial \mathcal{J}_\alpha} e^{i n_j q^j}\,,\nonumber \\
\frac{d \mathcal{J}_\alpha}{d \tau} &  =-i\epsilon \sum_{k \in \mathbb{Z}} k N_\alpha H_{k N_j} e^{i k N_j q^j}-i\epsilon \sum_{ n_j \in R}  n_\alpha H_{ n_j} e^{i  n_j q^j}
\end{align}
where $j\in\{r, \theta, \phi\}$ and $R$ is defined as the set of all non-resonant 3-tuples. In order to single out the effect of resonance in the presence of other fast-oscillatory non-resonant terms, we apply the technique of Near Identity Transformations as discussed in \cite{kevorkian1996near,vandeMeent:2013sza}, which is a transformation of the form
\begin{align}\label{eq:nit}
\tilde q^\alpha &= q^\alpha+\epsilon L^\alpha (\{q^\beta, \mathcal{J}_\beta\})+\mathcal{O}(\epsilon^2)\,,\nonumber \\
\tilde{\mathcal{J}}_\alpha &= \mathcal{J}_\alpha +\epsilon T_\alpha(\{q^\beta, \mathcal{J}_\beta\})+\mathcal{O}(\epsilon^2)\,,
\end{align}
with $L^\alpha, T_\alpha$ defined as
\begin{align}\label{eq:nitlt}
L^\alpha &= \sum_{n_j \in R} \frac{i}{n_j \Omega_j} \frac{\partial H_{n_j}}{\partial \mathcal{J}_\alpha} e^{i n_j q^j}\,, \nonumber \\
T_\alpha & = \sum_{n_j \in R} \frac{n_\alpha}{n_j \Omega_j} H_{n_j}e^{i n_j q^j}\,.
\end{align}

The dynamical variables $\tilde{q}^\alpha, \tilde{\mathcal{J}}_\alpha$ 
follow a set of equations of motion that are free of the ``contamination" of non-resonant terms:
\begin{align}\label{eq:eomtilde}
\frac{d \tilde{q}^\alpha}{d \tau} &  = \Omega_\alpha +\epsilon \sum_{k \in \mathbb{Z}} \frac{\partial H_{k N_j}}{\partial \mathcal{J}_\alpha} e^{i k N_j q^j}+\mathcal{O}(\epsilon^2)\,,\nonumber \\
\frac{d \tilde{\mathcal{J}}_\alpha}{d \tau} &  =-i\epsilon N_\alpha \sum_{k \in \mathbb{Z}} k  H_{k N_j} e^{i k N_j q^j}
+\mathcal{O}(\epsilon^2)\,.
\end{align}
In particular, the driving terms on the right-hand side of the equations always depend on a certain combination of angles: $N_j q^j$, which we shall define as the resonant angle $Q:= N_j q^j$. The rate of change of $Q$ is slow as compared to the frequencies of the non-resonant terms:
\begin{align}\label{eq:q}
\frac{d Q}{d \tau} = N_\alpha \Omega^\alpha + \epsilon \sum_{k \in \mathbb{Z}} N_\alpha \frac{\partial H_{k N_j}}{\partial \mathcal{J}_\alpha} e^{i k N_j q^j}
\end{align}
because $\Delta\Omega :=N_\alpha \Omega^\alpha +\epsilon N_\alpha \partial H_{(0,0,0)}/\partial \mathcal{J}_\alpha\approx 0$ based on the resonance assumption.
On the other hand, according to the second line of Eq.~\eqref{eq:eomtilde}, the change rate of each $\tilde{\mathcal{J}}_\alpha$ is proportional to each other: $\dot{\tilde{\mathcal{J}}}_r : \dot{\tilde{\mathcal{J}}}_\theta: \dot{\tilde{\mathcal{J}}}_\phi= N_r :N_\theta : N_\phi$. This means that we can write $\tilde{\mathcal{J}}_\alpha$ as
\begin{align}\label{eq:theta}
\tilde{\mathcal{J}}_\alpha = N_\alpha \Theta +C_\alpha
\end{align}
where $\Theta$ corresponds to the single dynamical ``action" and $C_\alpha$ are constants related to the choice of initial conditions.  The equation of motion for $\Theta$ is just
\begin{align}\label{eq:theta}
\frac{d \Theta}{d \tau} = -i\epsilon \sum_{k \in \mathbb{Z}} k  H_{k N_j} e^{i k N_j q^j}\,.
\end{align}

In fact, combing Eq.~\eqref{eq:q} and Eq.~\eqref{eq:theta}, we can view $\{Q, \Theta\}$ as a set of canonical variables for the effective Hamiltonian
\begin{align}\label{eq:he2}
\mathcal{H}_{\rm eff} &= \int \Delta\Omega d \Theta + 2 \epsilon\sum_{k \ge 1}{\rm Re}(H_{k {\bf N}}) \cos k Q \notag \\
&\qquad -2 \epsilon \sum_{k \ge 1}{\rm Im}(H_{k {\bf N}}) \sin k Q\,.
\end{align}
Because $\Theta$ is the only dynamical action after removing all non-resonant DOFs, we can expand $\Delta\Omega$ in power laws of $\Theta$ as
\begin{align}\label{eq:Delta_Omega}
\Delta\Omega = \alpha_0 +2\beta_0\Theta +\mathcal{O}(\Theta^2)
\end{align}
and dropping the higher-order terms. Substituting this into Eq.~\eqref{eq:he2}, we obtain the desired form of Eq.~\eqref{eq:he1}. In reality, for smooth $h$ we usually find that one of  harmonics dominate over the rest. Such harmonics $H_{\rm res}$ is likely much smaller then the amplitude of $h$ it self, as $h$ is generally dominated by the zeroth harmonic for many models of modified black hole spacetimes.
For the example perturbed Kerr spacetime we considered in this work, the perturbed Hamiltonian is dominated by the $\mathcal{H}_{n_r, n\theta=0}$ and $\mathcal{H}_{n_r, n\theta=\pm 2}$ components, where the former ones are non-resonant 
components. As a result,
for the 2/3 resonance [with $\mathbf{N}=(-3,2)\times k$] studied in the main text the dominant harmonics is $k_* = \pm 1$ and 
for the 1/2 harmonics  [with $\mathbf{N}=(-2,1)\times k$] discussed in the Appendix \ref{sec:app1} the dominant harmonics is $k_* =\pm 2$. 
In addition, we can always remove the $\sin Q$ term by adding/subtracting a  constant to the definition of $Q$. In the end, the effective Hamiltonian simplifies to ($H_{\rm res}=\pm 2 \epsilon |H_{\bf N}|$ depending on the sign of ${\rm Re}(H_{k {\bf N}})$):
\begin{align}\label{eq:eh}
\mathcal{H}_{\rm eff} = \alpha_0 \Theta +\beta_0 \Theta^2+H_{\rm res} \cos k_* Q\,,
\end{align}
which is governing the dynamics of the resonant DOF. Notice that there is additional ``gauge" freedom of $\alpha$ as one modified the definition of $\Theta$ by a constant:
\begin{align}
\Theta \rightarrow \Theta + c,\quad \alpha_0 \rightarrow \alpha_0 +2 \beta_0 c\,.
\end{align}
As a result, in practical implementations, we have chosen the minimal $\Theta$ of a trajectory to be zero to fix this gauge freedom. In this case, we have $\mathcal{H}_{\rm eff} = H_{\rm res}(\Theta=0)$ if the minimum locates at $Q=0$ ($\mathcal{H}_{\rm eff} = -H_{\rm res}(\Theta=0)$ if the minimum locates at $Q=\pm\pi$), and we empirically find that the dependence of $H_{\rm res}$ on $\Theta$ is weak (see Sec.~\ref{sec:nu}). Notice that in general we find the scaling of $H_{\rm res} $ to be $\epsilon^1$, the scaling of $\Theta, \alpha_0$ to be $\sqrt{\epsilon}$ and the scaling of $\beta_0$ to be one. If we rescale the variables using their $\epsilon$ scaling and adopt a ``slow-time" $\tilde{\tau} =\sqrt{\epsilon} \tau$, the resulting equations for $Q,\Theta$ will be $\epsilon$-free.
In particular,  the dynamical timescale for completing a ``cycle" in the $Q-\Theta$ plane is $\mathcal{O}(|\epsilon|^{-1/2})$ times the EMRI orbital timescale:
\begin{align}
t_{\rm res} \sim \mathcal{O} \left (\frac{1}{\omega \sqrt{H_{\rm res}}} \right ) \, ,
\end{align}
 where $\omega$ is the orbital angular frequency.

\subsection{Resonance  Crossing}\label{sec:rcross}

The effective Hamiltonian described by Eq.~\eqref{eq:eh} determines the resonant dynamics as the EMRI system is trapped in one of the resonant islands. This description will be explicitly verified by the numerical evolution of the system discussed in Sec.~\ref{sec:nu}. The form of the effective Hamiltonian, on the other hand, is very similar to the ones used to describe mean-motion resonances for multi-body planetary systems discussed in \cite{murray1999solar}. One difference is that $H_{\rm res}$ here does not necessarily follow the $\sqrt{\Theta}$ dependence usually assumed for planetary systems. Nevertheless, the form of the effective Hamiltonian naturally gives rise to two separate regimes in the phase space: the libration and the rotation regime as shown in Fig. \ref{fig:contour} in Sec.~\ref{sec:nu}.

Given $\alpha_0, \beta_0$ and $H_{\rm res}$, the system follows different trajectories in the $Q-\Theta$ phase space depending on the value of $\mathcal{H}_{\rm eff}$. 
Since $\mathcal{H}_{\rm eff} = H_{\rm res}(\Theta=0)$ and the $H_{\rm res}$ dependence on $\Theta$ is weak, we have
\begin{align}
\mathcal{H}_{\rm eff} = H_{\rm res} =\beta_0 \left ( \Theta+\frac{\alpha_0}{2\beta_0}\right )^2+H_{\rm res} \cos Q -\frac{\alpha^2_0}{4 \beta_0}
\end{align}
so that
\begin{align}
2H_{\rm res} \ge H_{\rm res}(1-\cos Q) \ge -\frac{\alpha^2_0}{4 \beta_0}\,.
\end{align}

The critical transition from a rotation orbit to a libration orbit happens when 
\begin{align}\label{eq:eec}
 H_{\rm res} = -\frac{\alpha_0^2}{8\beta_0}\ ,
\end{align}
 
The equality marks the condition for the critical trajectory entering/exiting a resonant island: $\alpha_c =\pm 2\sqrt{2}\sqrt{|\beta_0 H_{\rm res}|} \propto \sqrt{\epsilon}$. This transition may either be enabled by a parametric change of the effective Hamiltonian, or by an additional dissipative effect such as gravitational radiation reaction. Let us comment on these two effects individually.

First, let us focus on the parametric change of the effective Hamiltonian. With the radiation reaction turned on, the evolution equation for $\tilde{\mathcal{J}}_\alpha$ is modified 
\begin{align}\label{eq:g}
\frac{d \tilde{\mathcal{J}}_\alpha}{d \tau} &  =-i\epsilon N_\alpha \sum_{k \in \mathbb{Z}} k  H_{k N_j} e^{i k N_j q^j}+q \, G_\alpha \; ,
\end{align}
where $G_\alpha(\{\mathcal{J}_\beta\})$ is related to the orbit-averaged fluxes at infinity, which is approximately constant during the time of resonance. In addition, in practical implementations, we find that the $k=\pm 1, \pm 2$ harmonics usually dominate, and the $\partial H_{N_j}/\partial \mathcal{J}$ term in Eq.~\eqref{eq:q} is much smaller than the $\Delta\Omega$ term. As a result, the equations of motion may be simplified to (hereafter we focus on the $k_*=1$ case, where the generalization for other $k_*$ is straightforward)
\begin{align}\label{eq:sim}
\frac{d Q}{d \tau} &=\sum_\alpha a^\alpha \tilde{\mathcal{J}}_\alpha +b\,,\nonumber \\
\frac{d \tilde{\mathcal{J}}_\alpha}{d \tau} &  = -  N_{\alpha} H_{\rm res} \sin Q + q \, G_\alpha\,,
\end{align}
where $\sum_\alpha a^\alpha N_\alpha =\beta_0$ and $b$ is a constant. The parametric change of the effective Hamiltonian comes from the time-dependent shift of $\mathcal{J}_\beta$ with $N_\beta =0$. For  a $r-\theta$ resonance (as discussed in Sec.~\ref{sec:nu}), $N_\phi, N_t$ are zero so the parametric shift is induced by $\mathcal{J}_t, \mathcal{J}_\phi$ (the energy and angular momentum).   
Since $G_t, G_\phi$ induce the time-dependent modulation of the conserved quantity $\mathcal{J}_t, \mathcal{J}_\phi$, the value of $\alpha_0, \beta_0, H_{\rm res}$ will in turn shift in time as they are all functions of $\mathcal{J}_t, \mathcal{J}_\phi$. Such shifts can be viewed as a parametric change of the effective Hamiltonian. If the parametric shift timescale through the resonance regime ($\mathcal{O}(1)$ change of resonant angle $Q$  due to radiation reaction) $t_{\rm rr}\propto 1/\sqrt{\Delta \dot{\Omega}} \propto \omega^{-1}/\sqrt{q}$  is longer than $t_{\rm res}$, the change is adiabatic such that the action of the effective Hamiltonian is invariant, i.e., 
\begin{align}
\mathcal{I} =\int \Theta d Q
\end{align}
is constant. According to the phase-space analysis in \cite[Chap.~8]{murray1999solar}, the adiabatic capture into a resonance is only possible for ``converging" evolution, i.e., the ratio of two frequencies evolves towards one or equivalently $\alpha$ (assuming $\beta$ is negative) evolves from positive values to negative values. In the $r-\theta$ resonance considered in Sec.~\ref{sec:nu}, $\alpha$ generally evolves from negative values to positive values with the corresponding frequency ratio differing more in time. Using the notation of the rotation number $\nu$, under the influence of radiation reaction, we find that $\nu$ in general decreases across resonances, e.g., evolving from $>2/3$ to $<2/3$ across the $2/3$ resonance.
As a result, adiabatic parametric resonance capture is unlikely. This argument should apply for generic choice of $h$.

The second transition mechanism comes from the evolution of $\mathcal{J}_r, \mathcal{J}_\theta$ in Eq.~\eqref{eq:sim}. The equation of motion is equivalent to one-dimensional motion with total effective energy 
\begin{align}\label{eq:k}
\frac{1}{2}\left ( \frac{d Q}{d \tau}\right )^2+\beta_0 H_{\rm res} \cos Q +q \sum_\alpha a_\alpha G_\alpha Q =K\,. 
\end{align}
As shown in \cite{vandeMeent:2013sza}, this type of evolution will not lead to the capture into a resonant island unless higher order radiation reaction is considered, in which case the capture condition is still extremely fine-tuned. In summary, the dissipation in $\mathcal{J}_r, \mathcal{J}_\theta$ also does not lead to resonance trapping in the adiabatic regime.

Once the adiabatic approximation breaks down, it is possible to have trajectories across the resonant islands. The island crossing time can be estimated as
\begin{align}
t_{\rm cross} \sim \frac{\alpha_c}{\dot{\alpha_0}} \propto \frac{\sqrt{H_{\rm res}}}{q \omega}
\end{align} 
where $\omega$ is the orbital frequency, $\dot{\alpha_0} \propto q$ is a parametric shift of $\alpha_0$ driven by the radiation reaction. Therefore, the adiabatic approximation breaks down if $t_{\rm cross} < t_{\rm res}$, or $H_{\rm res} < q$ (apart from numerical coefficients). Notice that we write $H_{\rm res}$ instead of $h$ here because the harmonics amplitude $H_{\rm res}$ can be much smaller than the perturbation amplitude $h$. This point is also seen in the numerical example studied in Sec.~\ref{sec:nu}.

To summarize:
in the adiabatic limit, radiation dissipation does not drive a near-resonance orbit onto resonance 
due to the existence of the adiabatic invariant $\mathcal{I}$. It seems no obvious pathway for a near-resonance orbit 
to cross the resonance in this limit. But as we will see later, the effective (1-DOF) Hamiltonian description we have been using breaks down for chaotic transitional orbits, which are possible to jump from one branch to the other branch with opposite sign of $dQ/d\tau$ even without radiation dissipation. Therefore, chaotic transitional orbits are the pathway of crossing resonances without crossing the resonance islands  in the adiabatic limit, while crossing the resonance islands is a more general pathway for crossing resonances when the adiabatic approximation breaks down.

\subsection{The Emergence of Chaos}

It is interesting to discuss the emergence of chaos in the Hamiltonian point of view, especially outside the resonant islands. According to the derivation in Sec.~\ref{sec:subeff}, an Near Identity Transformation can be applied to remove all the fast-varying degrees of freedom, so that the resonant effective Hamiltonian only has one degree of freedom left. If this description is complete, then no chaos should appear for the resonant degree of freedom, which clearly contradicts the numerical observations.

One likely explanation is that when we perform the Near Identity Transformation in Eq.~\ref{eq:niteom}, \ref{eq:nit} and \ref{eq:nitlt}, we have neglected terms with higher order than $\mathcal{O}(\epsilon)$. These terms may not necessarily come from nonlinear $h$. For example, let us consider a non-resonant angle $Q+q_\theta$ that may appear in some of the terms in Eq.~\ref{eq:niteom}. In the process of Near Identity Transformation we have included the time dependence of $q_\theta$ as $\Omega_\theta \tau$, but have neglected the time dependence of $Q$ because it evolves on the slow time $\propto \sqrt{\epsilon}$. This assumption is valid for trajectories within the resonant islands, where the value of $Q$ is bounded in a region less than $2\pi$. However, outside the resonant islands when $Q$ starts to rotate, this part of phase produces secular phase errors for evolution longer than $1/\sqrt{\epsilon}$ times the orbital timescale.

In Chapter 9.5 of \cite{murray1999solar}, it is shown that if additional time-dependent terms is added to the Hamiltonian of a simple pendulum (similar to the effective Hamiltonian here):
\begin{align}
\mathcal{H}_{\rm eff} \rightarrow \mathcal{H}_{\rm eff}+H_{\rm non-res}\sum_k \cos (Q+k \omega \tau)
\end{align}
where $H_{\rm non-res}, \omega$ are constants, chaos appears. The emergence of chaos is studied using the ``Standard Map" method in \cite{murray1999solar}. Notice these additional terms look at the non-resonant terms we remove using the Near Identity Transformation. Therefore it is plausible that the interplay between the non-resonant terms and the resonant terms in Eq.~\ref{eq:niteom} gives rise to the chaotic region outside the resonant islands. In other words, it is beneficial to keep both resonant and no-resonant terms in modelling the trajectories in the chaotic regime, i.e., using $q^\alpha, \mathcal{J}^\alpha$. On the other hand, although the single DOF effective Hamiltonian cannot fully describe the transitional chaotic orbits, it is still useful in this regime because the non-chaotic trajectories predicted by the effective Hamiltonian can be still viewed as temporary orbits in the chaotic zone, especially in the region away from the bifurcation points. The transitional chaotic orbits may be viewed as collections of non-chaotic orbits in the $Q-\Theta$ plane with occasional transition from one to the other.

\section{Numerical Evolution}\label{sec:nu}

In this Section, we will use the spinning BH solution in quadratic gravity~\cite{Maselli:2015tta,Dong:2021yss} as an example perturbed Kerr spacetime
for clarifying the resonance crossing process. This spacetime has been numerically studied before \cite{Cardenas-Avendano:2018ocb} which is convenient for us to validate and compare the part of our results in physical coordinates. While being obtained from a specific quadratic gravity theory, the results presented in this Section should apply to various perturbed Kerr metrics in general.

\subsection{Quadratic gravity}

For the purpose of numerical evolution, we have focused on a particular quadratic gravity theory, namely Einstein-scalar-Gauss-Bonnet (EsGB) gravity. Quadratic gravity usually arises as an effective field theory for the low energy limit of some quantum gravity theories~\cite{Kanti:1995vq,Moura:2006pz}. In addition, EsGB is the only theory quadratic in the curvature that leads to second-order field equations for any coupling, ensuring that the theory is ghost-free~\cite{Woodard:2006nt}.

It introduces an additional scalar field $\vartheta$ coupled to the metric $g_{\mu\nu}$ through the action
\begin{align}
    S = \int\mathrm{d}^{4}x\,\sqrt{-g}\left[\kappa R + \alpha_{\mathrm{GB}}\,\vartheta\,\mathcal{G} - \frac{1}{2}\left(\nabla_{\mu}\vartheta\,\nabla^{\mu}\vartheta + 2V(\vartheta)\right) + \mathcal{L}_{\rm mat}\right],
\end{align}
where $\kappa=(16\pi G)^{-1}$, $g$ is the determinant of the metric, $R$ is the Ricci scalar, $\mathcal{G}\equiv R^2 - 4 R_{\mu\nu}R^{\mu\nu} + R_{\mu\nu\rho\sigma}R^{\mu\nu\rho\sigma}$ is the Gauss-Bonnet invariant, $V(\vartheta)$ is a potential that will be set to zero in the following and $\mathcal{L}_{\rm mat}$ is the Lagrangian density describing the matter fields. The coupling constant $\alpha_{\mathrm{GB}}$ has the dimension of $[\mathrm{length}]^2$, so we also define the following dimensionless constant 
\begin{equation}
    \zeta \equiv \frac{\alpha_{\mathrm{GB}}^2}{\kappa M^4},
\end{equation}
where $M$ is the mass of the black hole.
The set of field equations for the theory are 
\begin{align}
&    G_{\mu\nu} + \frac{\alpha_{\mathrm{GB}}}{\kappa} \mathcal{D}_{\mu\nu}^{(\vartheta)} = \frac{1}{2\kappa} \left(T_{\mu\nu}^{(\mathrm{mat})} + T_{\mu\nu}^{(\vartheta)} \right)\,, \\
&    \square\vartheta = \alpha_{\mathrm{GB}}\,\mathcal{G}\,,
\end{align}
where $T_{\mu\nu}^{(\mathrm{mat})}$ and $T_{\mu\nu}^{(\vartheta)}$ are respectively the matter and scalar field stress-energy tensors with
\begin{align}
    T_{\mu\nu}^{(\vartheta)} = \frac{1}{2}g_{\mu\nu}\left(\nabla_{\rho}\vartheta\,\nabla^{\rho}\vartheta - 2V(\vartheta)\right) - \nabla_{\mu}\vartheta\,\nabla_{\nu}\vartheta \,,
\end{align}
and
\begin{align}\nn
    \mathcal{D}_{\mu\nu}^{(\vartheta)} = & -2 R\nabla_{\mu}\nabla_{\nu}\vartheta + 2\left(g_{\mu\nu}R-2R_{\mu\nu}\right)\nabla_{\rho}\vartheta\,\nabla^{\rho}\vartheta \\
    & +8R_{\rho(\mu}\nabla_{\nu)}\nabla^{\rho}\vartheta - 4g_{\mu\nu}R^{\rho\sigma}\nabla_{\rho}\nabla_{\sigma}\vartheta +4 R_{\mu\nu\rho\sigma}\nabla^{\rho}\nabla^{\sigma}\vartheta\,.
\end{align}

This theory has been extensively studied in the past and solutions have been obtained both numerically~\cite{Pani:2009wy,Kleihaus:2014lba} and analytically~\cite{Yunes:2011we,Ayzenberg:2014aka,Maselli:2015tta}.  Spherically symmetric solutions are known to lead to integrable orbits, so here we will only focus on stationary and axisymmetric solutions. As no exact closed-form solutions are known for rotating BHs, analytical solutions are obtained as an expansion in the small coupling and for small rotation. This is the path we will follow in the following.

These metrics were obtained following two independent approximation schemes, one for small coupling $\zeta\ll 1$ and the other one for small spin $\chi\equiv \frac{a}{M}\ll 1$, with $a$ the dimensional spin parameter. Then, the solution was resummed in order to interpret the solution as a perturbation of the Kerr metric~\cite{Dong:2021yss}
\begin{align}
    g_{\mu\nu} = g_{\mu\nu}^{(\mathrm{Kerr})} + \sum_{l,m} \zeta'^l\,\alpha'^m\,\delta g_{\mu\nu}^{(l,m)}\,,
\end{align}
where $\alpha'$ and $\zeta'$ are bookkeeping parameters and $g_{\mu\nu}^{(\mathrm{Kerr})}$ is the usual Kerr metric in Boyer-Lindquist coordinates
$\left(t, r, \theta, \phi\right)$
\begin{align}\label{eq:kerr} \nn
    &g_{tt}^{(\mathrm{Kerr})} = -\left(1-\frac{2Mr}{\Sigma}\right),\quad g_{t\phi}^{(\mathrm{Kerr})} = -\frac{2Mar\sin^2\theta}{\Sigma},\\ \nn
    &g_{rr}^{(\mathrm{Kerr})} = \frac{\Sigma}{\Delta},\quad g_{\theta\theta}^{(\mathrm{Kerr})} = \Sigma,\\
    & g_{\phi\phi}^{(\mathrm{Kerr})} = \left(r^2+a^2+\frac{2Ma^2r\sin^2\theta}{\Sigma}\right)\sin^2\theta\,,
\end{align}
where $\Sigma \equiv r^2+a^2\cos^2\theta$ and $\Delta \equiv r^2-2Mr+a^2$.
As we treat quadratic gravity as an effective field theory, we will only need the linear order in the coupling $l=1$ and consider the set of perturbations $\delta g_{\mu\nu}^{(1,m)}$. As an example, we give the explicit expressions of the perturbation up to linear order in spin:
\begin{align}\nn
    &\delta g^{(1,0)}_{tt} = -\frac{\zeta M^3}{3r^2} \left[1+\frac{26M}{r}+\frac{66M^2}{5r^2}+\frac{96M^3}{5r^3}+\frac{80M^4}{r^4}\right], \\ \nn
    &\delta g^{(1,0)}_{rr} = -\frac{\zeta M^2}{f^2r^2} \left[ 1+\frac{m}{r}+\frac{52m^2}{3r^2}+\frac{2M^3}{r^3}+\frac{16m^4}{5r^4}-\frac{368m^5}{3r^5} \right], \\ \nn
    &\delta g^{(1,1)}_{t\phi} = \frac{3}{5}\zeta M\chi\,\frac{M^3\sin^2\theta}{r^3} \left[ 1+\frac{140M}{9r}+\frac{10M^2}{r^2}\right. \\
    &\left.\qquad\qquad\qquad\qquad\qquad\qquad\qquad +\frac{16M^3}{r^3}-\frac{400M^4}{9r^4} \right]\,,
\end{align}
with $f\equiv 1-\frac{2M}{r}$ and all other components vanish. In the following, we will use the solution up to quintic order in spin but as the expressions are quite lengthy, we refer to Ref.~\cite{Dong:2021yss} for the full expressions.

\subsection{Evolution in physical coordinates $(x^\mu, p_\mu)$}
We consider the spinning BH solution of the quadratic gravity as an example of perturbed Kerr spacetime, 
where the perturbed spacetime is still stationary and axisymmetric. 
For a test particle moving in the perturbed Kerr spacetime along a geodesic, 
the total Hamiltonian $H$, the energy $E$ and the angular momentum $L$ are still conserved, 
but the Carter constant $C$ is not. 

\begin{figure*}
\includegraphics[scale=0.9]{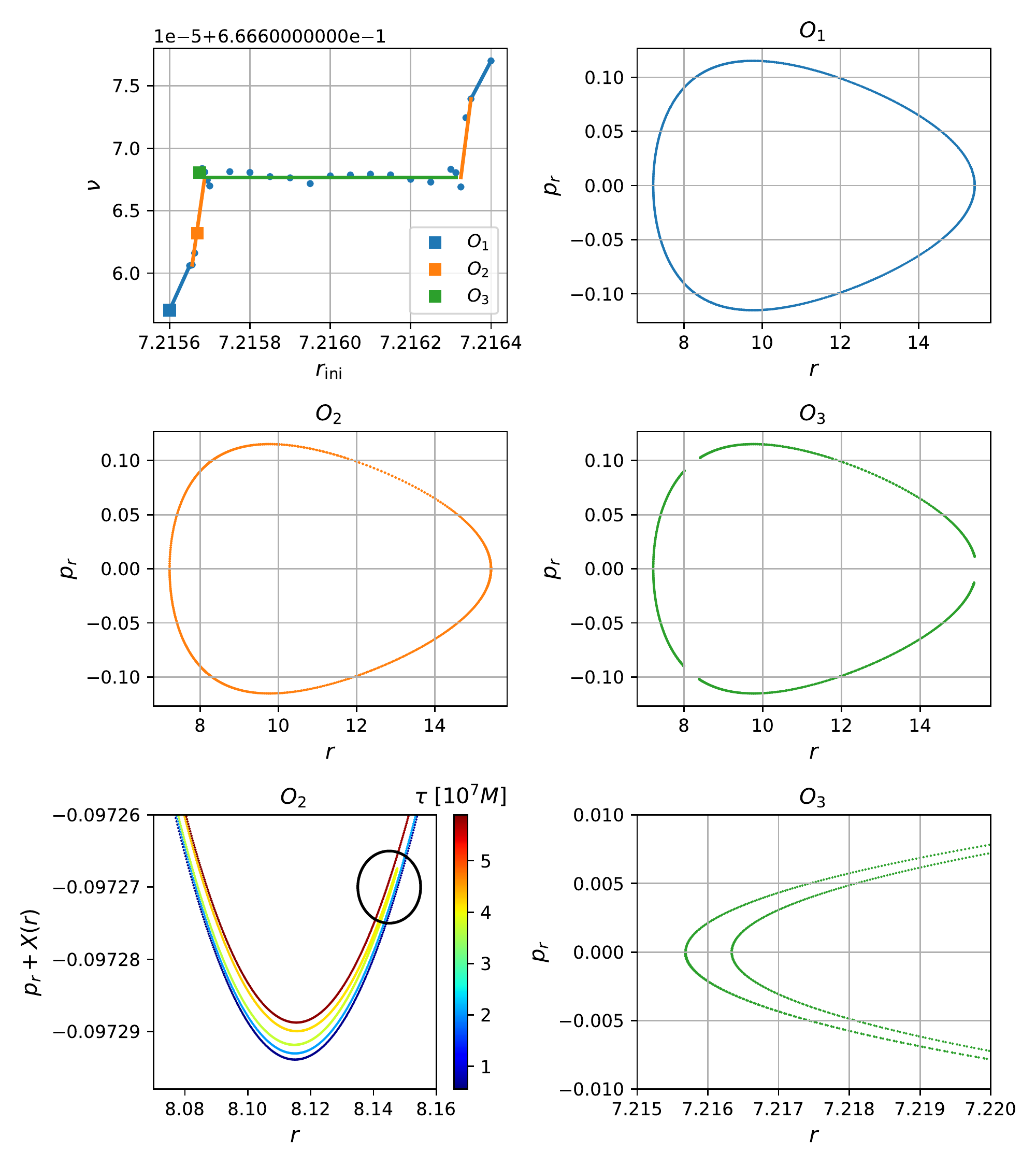}
\caption{\label{fig:Poincare} 
Near-resonance orbits in the perturbed Kerr spacetime with parameter $(a, \zeta)=(0.2, 0.002)$.
Upper left panel: the rotation numbers of near-resonance orbits, with $(E, L)=(0.96, 3.5)$ and initial conditions $(\theta, p_r)_{\rm ini}=(\pi/2, 0)$, where the dots are the numerical results and the straight lines of three different colors
denote three different kinds of orbits: regular orbits in blue, chaotic transitional orbits in orange and on-resonance orbits in green, and the three square dots are three representative orbits  $O_{1/2/3}$ (blue/orange/green) with initial radius $r_{\rm ini}=(7.2156, 7.21566875, 7.215675)$, respectively. Upper right/middle left/middle right panels: Poincar\'e maps $(r, p_r)|_{\theta=\pi/2}$ of orbits $O_{1/2/3}$, respectively. 
Lower left panel: zoomed-in Poincar\'e map of  orbit $O_2$, 
and $X(r)=3.38\times 10^{-2}\times(r-8.2)$, where a turn-back shows up around $r=8.15$ at $\tau\approx4\times10^7$.
Lower right panel: zoomed-in Poincar\'e map of orbit $O_3$, where an island clearly shows up.}
\end{figure*}

In this work, we choose a perturbed Kerr spacetime with BH spin $a=0.2$ and the quadratic gravity coefficiency $\zeta=0.002$ as a fiducial example. To get an intuition for the chaotic behavior of near-resonance orbits in a perturbed Kerr spacetime, we consider a series of geodesics with total Hamiltonian $H=-1/2 M^2$ (or equivalently the particle rest mass $\mu=1$), energy $E=0.96 M$, angular momentum $L=3.5 M^2$, and initial condition
$(r, \theta, p_r)_{\rm ini}=(r_{\rm ini}, \pi/2, 0)$ --- $(p_\theta)_{\rm ini}$ is then automatically determined and solved for numerically.
Given the initial conditions, we evolve the system according to the Hamiltonian equations
\be 
\dot x^A = \frac{\partial H(x^\mu, p_\mu)}{\partial p_A}\ ,\quad \dot p_A = -\frac{\partial H(x^\mu, p_\mu)}{\partial x^A} \\ 
\ee 
from $\tau_{\rm ini}=0$ to $\tau_{\rm end}=6\times10^7 M$, with the index $A$ running over $(r,\theta)$.

For each orbit, we calculate the rotation number 
\be 
\nu : = \frac{\braket{\Omega^r}}{\braket{\Omega^\theta}}\ ,
\ee 
which is the ratio of average frequency in the $r$ direction and that in the  $\theta$ direction.
Numerically it can be computed as $\nu \approx N_{\dot r=0}/N_{\theta=\pi/2}$, where $N_{\dot r=0}$ is the number of times the orbit passes the pericenter and apocenter, and $N_{\theta=\pi/2}$ is the number of times the orbit passes the equator in the same time interval. In the top left panel of Fig.~\ref{fig:Poincare}, we show the rotation numbers of the orbits (labelled by $r_{\rm ini}$) that are close to the $2/3$ resonance. 
The near-resonance orbits can be classified into three categories:  orbits in the rotation regime (blue horizontal line), chaotic transitional orbits (orange line) and  orbits in the libration regime (green line).

Similar features in the rotation curve have been found in previous studies on various non-Kerr metrics (see e.g., \cite
{Deich:2022vna,Bronicki:2022eqa,Destounis:2020kss,Destounis:2021mqv,Destounis:2021rko,Destounis:2023gpw,Lukes-Gerakopoulos:2017jub,
Apostolatos:2009vu,Lukes-Gerakopoulos:2010ipp,Cardenas-Avendano:2018ocb,Zelenka2020}). These features (including both the plateau and jumps in the rotation curve) were believed to be the signatures of chaos.  However, as we will show later, only the steep part of the rotation curve (orange line) correspond to chaotic orbits, while the remaining parts correspond to non-chaotic orbits: either on-resonance non-chaotic orbits (the plateau part, blue line) or the near-resonance non-chaotic orbits (the mildly changing part, green line).

To better understand the orbits in each category, we also plot the Poincar\'e map $(r, p_r)_{\theta=\pi/2}$ of three representative orbits $O_{1/2/3}$ with $r_{\rm ini}=(7.2156, 7.21566875, 7.215675) M$ in Fig.~\ref{fig:Poincare}. 
In the upper right/ middle left/ middle right panels, we show the Poincar\'e maps of these three orbits $O_{1,2,3}$,
where $O_1$ and $O_2$ are similar simple closed curves, while $O_3$ consists of 3 disconnected pieces. 
In the lower left panel, we show a zoomed-in version of the $O_2$ map around 
$r=8 M$. An interesting feature 
is that the curve turns back around $r=8.15 M$ at $\tau\approx4\times10^7 M$, i.e., the map $(r, p_r)$ moves in the anti-clockwise direction before this time and in the clockwise direction afterwards. This turning back behavior is a key signature of transitional chaotic orbits, which allows orbits to switch from one branch (i.e., the ``anti-clockwise" branch) to the other (i.e., the ``clockwise" branch) without crossing the resonant islands. This transition is further discussed using the language of action-angle variables in Sec.~\ref{sec:aa} (see Fig.~\ref{fig:phase_O2}).   
 In the lower right panel, we show a similarly zoomed-in version of the $O_3$ map around $r=r_{\rm ini}$, where the map is in fact an ``island" instead of a simple curve.

Though we have seen rich phenomena of near-resonance EMRI dynamics in the physical coordinates, no simple picture or unified description of these phenomena is available. In order to obtain a systematic and quantitative description of the underlying physics, working with action-angle variables is generally useful.

\subsection{Evolution in action-angle variables $(\mathcal{J}_\alpha, q^\alpha)$ and in resonant variables $(\Theta, Q)$}\label{sec:aa}

\begin{figure*}
\includegraphics[scale=0.8]{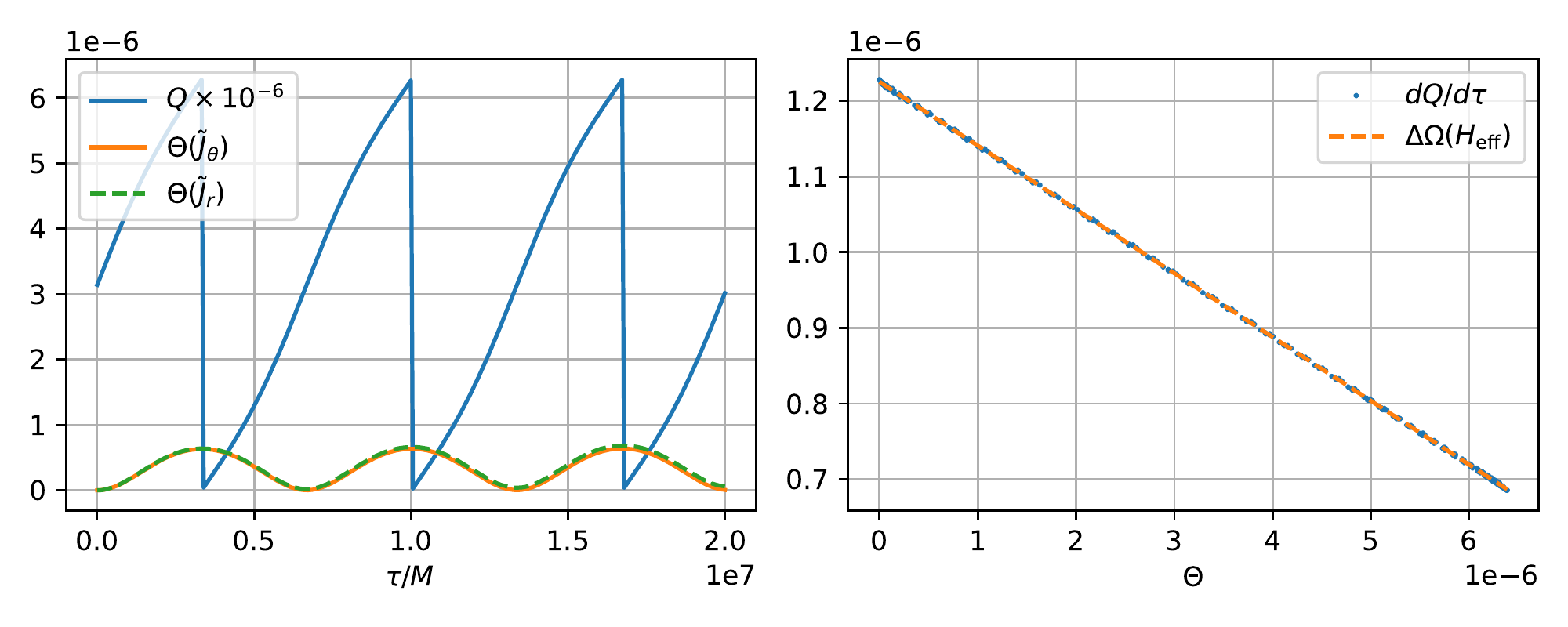}
\caption{\label{fig:eff_O1} Left panel: evolution of near-resonance orbit $O_1$ in terms of resonant angle $Q$ and the conjugate momentum $\Theta$. Right panel: the resonant angular velocity $\dot Q(\Theta)$, where the dots are the numerical derivative of $Q(\tau)$ and the straight line is $\Delta\Omega(\Theta)=\alpha_0+2\beta_0\Theta$ from Eq.~(\ref{eq:Delta_Omega}).} 
\end{figure*}

\begin{figure*}
\includegraphics[scale=0.6]{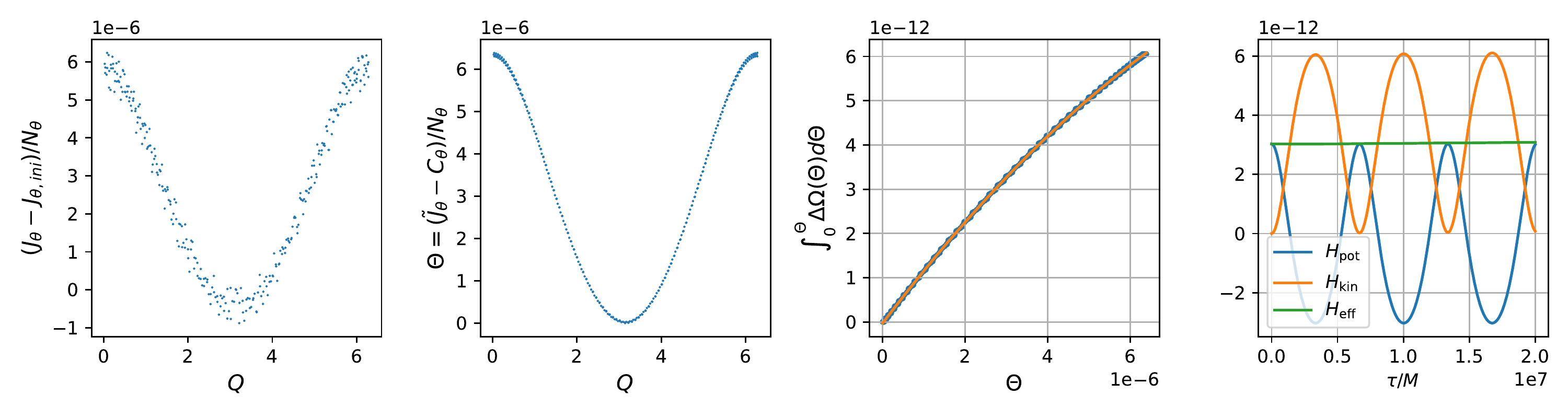}
\caption{\label{fig:phase_O1} First panel: phase diagram $J_\theta-Q$ of orbit $O_1$. 
Second panel: phase diagram $\Theta(\tilde{J_\theta})-Q$ of orbit $O_1$.
Third panel: the effective kinetic energy $H_{\rm kin}:=\int_0^\Theta \Delta\Omega(\Theta) d\Theta$
as a function of momentum $\Theta$, where the dots are the numerical results of 
$H_{\rm pot}(\Theta=0)-H_{\rm pot}$ and the solid line is a quadratic fitting $\int_0^\Theta \Delta\Omega(\Theta) d\Theta = c_0+\alpha_0\Theta + \beta_0\Theta^2$. Fourth panel: numerical result of potential energy $H_{\rm pot}(\tau):= \sum H_{ k} e^{ikQ}$,
fitting result of kinetic energy $H_{\rm kin}(\tau):=c_0+\alpha_0\Theta + \beta_0\Theta^2$, and the summation of the two $H_{\rm eff}(\tau):=H_{\rm pot}(\tau)+H_{\rm kin}(\tau)$.}
\end{figure*}

\begin{figure*}
\includegraphics[scale=0.6]{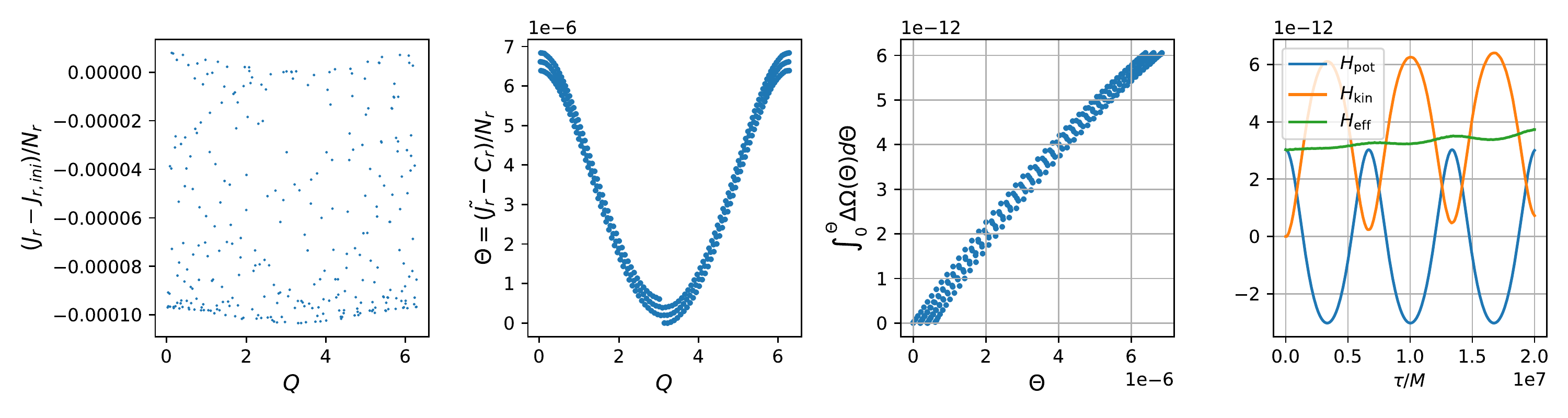}
\caption{\label{fig:phase_O1b} First panel: phase diagram $J_r-Q$ of orbit $O_1$. 
Second panel: phase diagram $\Theta(\tilde{J_r})-Q$ of orbit $O_1$.
Third panel: the effective kinetic energy $H_{\rm kin}:=\int_0^\Theta \Delta\Omega(\Theta) d\Theta$
as a function of momentum $\Theta$, where the dots are the numerical results of 
$H_{\rm pot}(\Theta=0)-H_{\rm pot}$. Fourth panel: numerical result of potential energy $H_{\rm pot}(\tau):= \sum H_{k} e^{ikQ}$, fitting result of kinetic energy $H_{\rm pot}(\tau):=c_0+\alpha_0\Theta + \beta_0\Theta^2$, and the summation of the two $H_{\rm eff}(\tau):=H_{\rm pot}(\tau)+H_{\rm kin}(\tau)$.}
\end{figure*}

In Sec.~\ref{sec:Heff}, we have presented an (effective) Hamiltonian description of the resonant DOF in the resonance regime. In going from the full Hamiltonian to the effective Hamiltonian, all the non-resonant degrees of freedom have been removed by applying a NIT. This is valid because the non-resonant degrees of freedom act as fluctuations to the resonant dynamics, which evolves on the ``slow-time": $\tilde{\tau} =\sqrt{\epsilon}\tau$. When the resonant motion is switched to the rotational regime, the interplay between resonant and non-resonant degrees of freedom gives rise to the transitional chaotic orbits that cannot be fully characterized by the effective Hamiltonian. Nevertheless, the Hamiltonian description offers a general mathematical framework to analyze the resonance phenomena. We shall transform the orbit information obtained in the previous section in the physical coordinates to the action-angle coordinates and numerically verify  the formalism discussed in Sec.~\ref{sec:Heff}. The technical tools for making such transformations are explicitly given in Appendix~\ref{sec:app2} and \ref{sec:app3}. To the best of our knowledge, this is also the first time that EMRI evolutions (in the perturbed Kerr spacetime without radiation reaction included) are shown with action-angle variables.

We start by plotting the resonant angle and actions for the sequence of orbits shown in Fig.~\ref{fig:Poincare} evolved using physical coordinates. In the left panel of Fig.~\ref{fig:eff_O1}, we show the evolution of the resonant angle $Q(\tau)$ and the conjugate momentum 
$\Theta(\tau)$ of orbit $O_1$, where $Q(\tau)$ monotonically increases, which is consistent with $\nu < 2/3$ as 
\be 
\dot Q = -3 \dot{\tilde{q}}^r+2 \dot{\tilde{q}}^\theta = -3\braket{\Omega^r}+2\braket{\Omega^\theta}
=2\braket{\Omega^\theta} \left(1-\frac{3}{2}\nu\right) \ .
\ee 
The plot shows the resonant momentum $\Theta$ obtained from $\tilde{\mathcal{J}}_\theta$ and $\tilde{\mathcal{J}}_r$, the results are consistent with each other, as a direct verification for Eq.~\eqref{eq:theta}.

By interpolating $Q(\tau)$, we numerically obtain the derivative $\dot Q(\tau)$ and show $\dot Q(\Theta)$ in the right panel,
which is almost a perfect straight line. 
As a consistency check, we can also infer the resonant angular velocity $\Delta\Omega(\Theta)$ via the effective Hamiltonian.
Writing the effective Hamiltonian as the kinetic energy and the potential energy
\be 
H_{\rm eff} := H_{\rm kin} + H_{\rm pot} = \int_0^\Theta \Delta\Omega(\Theta) d\Theta + \sum_k H_{k\mathbf{N}} e^{ikQ}\ ,
\ee 
it is straightforward to see $H_{\rm eff}=H_{\rm pot}(\Theta=0)$, therefore we have $H_{\rm kin}=H_{\rm pot}(\Theta=0)-H_{\rm pot}(\Theta)$. We numerically fit $H_{\rm kin}(\Theta)$ with $c_0+\alpha_0\Theta+\beta_0\Theta^2$, and obtain the linear relation $\Delta\Omega(\Theta)=\alpha_0+2\beta_0\Theta$, which perfectly matches the numerical results of $\dot Q(\Theta)$
as shown in the right panel of Fig.~\ref{fig:eff_O1}.

Let us now look at the phase diagram for the resonant DOF. In the first panel of Fig.~\ref{fig:phase_O1}, we show the  diagram $J_\theta-Q$ of orbit $O_1$, which is quite noisy because $J_\theta$ contains both resonant and non-resonant degrees of freedom.
After the NIT, the non-resonant degrees of freedom are removed and we obtain the plots for $\tilde J_\theta-Q$, which clearly shows that the near-resonance orbit $O_1$ is effectively a system of single DOF. In addition, we see that the orbit $O_1$ is in the rotation regime because the resonant angle 
$Q$ covers the full range of $[0, 2\pi]$.
In the third panel, we show the kinetic part $H_{\rm kin}$ of the effective Hamiltonian and a quadratic fitting to it.
In the fourth panel, we show the numerical potential energy $H_{\rm pot}$, the fitted kinetic energy $H_{\rm kin}$ and the total energy $H_{\rm eff}=H_{\rm kin}+H_{\rm pot}$, which is a constant as expected.

Fig.~\ref{fig:phase_O1b} is similar to Fig.~\ref{fig:phase_O1} except we consider the action $J_r$ instead of $J_\theta$.
In the left panel, we barely see any correlation between the action $J_r$ and the resonant angle $Q$ because the non-resonant degrees of freedom dominates the variations of $J_r$.  The curve 
$\Theta(\tilde{J}_r)$ is recovered via the NIT, which reduces the variations by more than one order of magnitude (see Appendix \ref{sec:app2} and \ref{sec:app3} for the details regarding the implementation of the NIT), though dispersion remains due to the presence of relatively large numerical errors in the NIT procedure.
As a result, we find a similar dispersion in the kinetic energy $H_{\rm kin}(\Theta)$ and consequently the conservation of the effective Hamiltonian $H_{\rm eff}(\tau)$ is not as good as in the $J_\theta$ case. In the following analyses, 
we will use the higher-precision $\Theta(\tilde{J}_\theta)$.

\begin{figure*}
\includegraphics[scale=0.8]{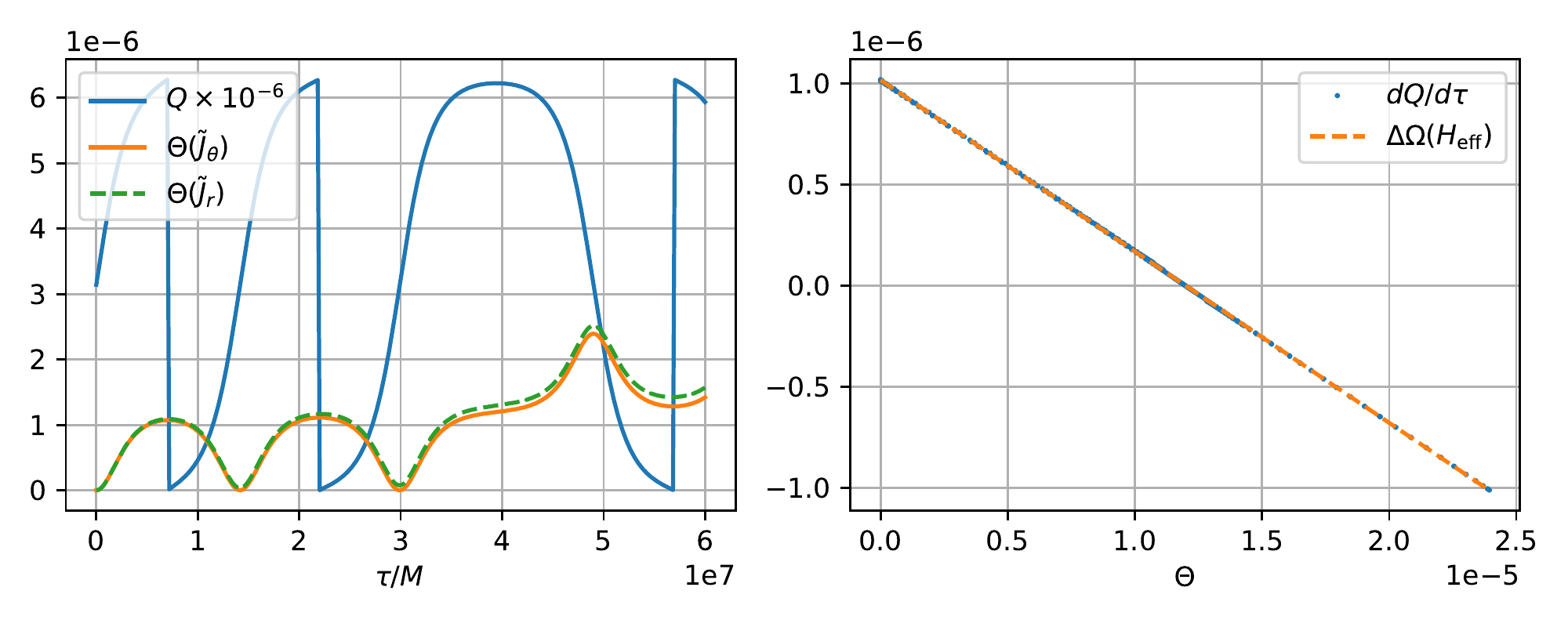}
\caption{\label{fig:eff_O2} Similar to Fig.~\ref{fig:eff_O1} except for orbit $O_2$. In the left panel, a turn-back in $Q$ and a jump in $\Theta$ appear around $\tau\approx4\times10^7 M$.} 
\end{figure*}

\begin{figure*}
\includegraphics[scale=0.6]{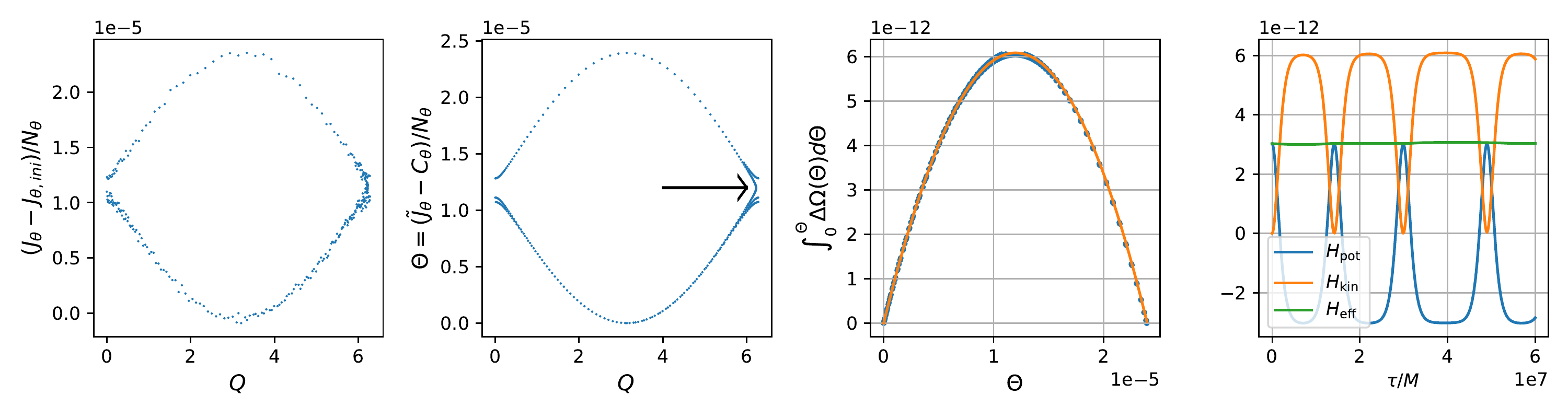}
\caption{\label{fig:phase_O2} Similar to Fig.~\ref{fig:eff_O1} except for orbit $O_2$. In the second panel, two branches show up in the $Q-\Theta$ phase diagram and the transition from one branch to another 
occurs at $Q\approx 2\pi$.}
\end{figure*}

To illustrate chaotic orbits, we perform a similar analysis for orbit $O_2$ and show the results in Figs.~\ref{fig:eff_O2} and \ref{fig:phase_O2}.
In the left panel of Fig.~\ref{fig:eff_O2}, we see that this orbit consists of two branches, a $Q$ increasing branch (the ``counter-clockwise" branch in Fig.~\ref{fig:Poincare})
and a $Q$ decreasing branch (the ``clockwise" branch in Fig.~\ref{fig:Poincare}), and a transition occurs around $\tau=4\times10^7 M$. This transition is also marked by a jump in momentum $\Theta$, which in turn induces an extra phase shift in the waveform as we will discuss in Sec.~\ref{sec:impact}. As shown in the right panel, $\Delta\Omega(\Theta)$ on the two different branches follows the same linear relation, which indicates the existence of a symmetry between the two branches. This symmetry is revealed more explicitly in the second panel of Fig.~\ref{fig:phase_O2}, where the $\Theta-Q$ diagram consists of two symmetric branches that are connected by a sharp transition around $Q=2\pi$ (the bifurcation point indicated by the arrow).

\begin{figure*}
\includegraphics[scale=0.8]{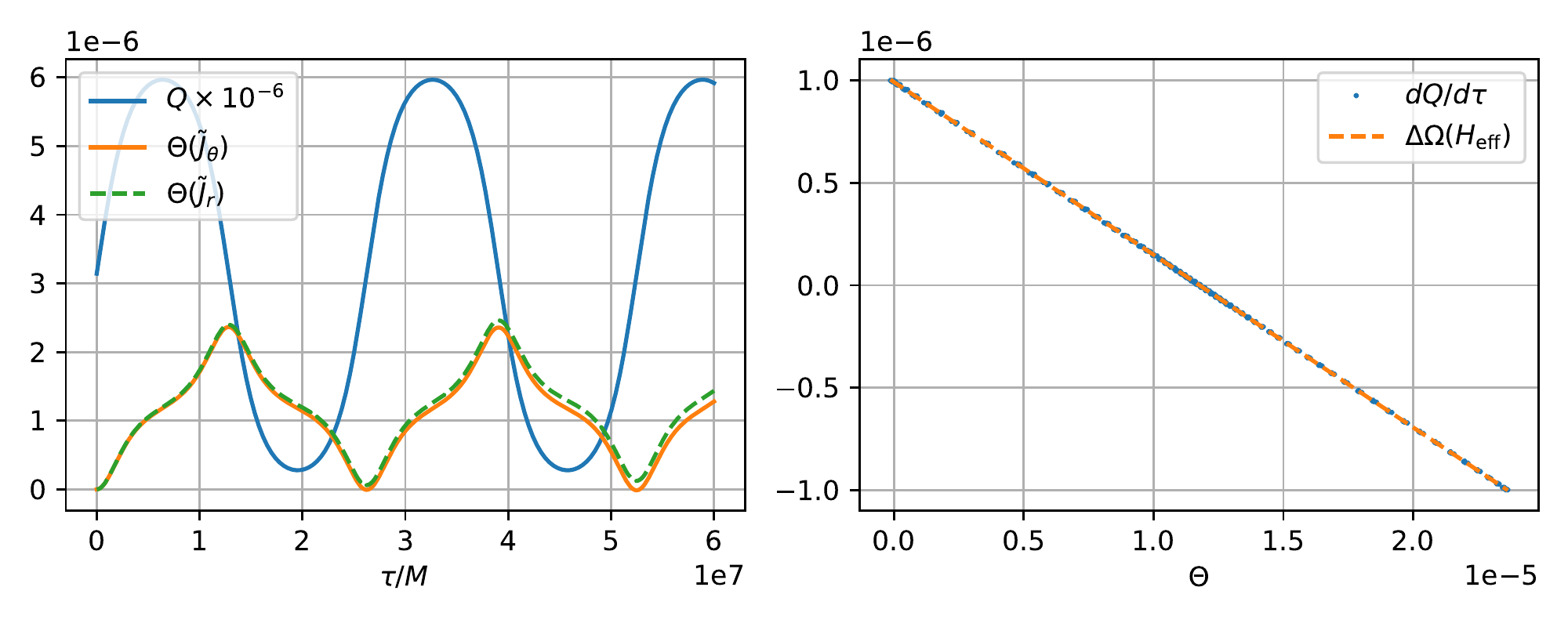}
\caption{\label{fig:eff_O3} Similar to Fig.~\ref{fig:eff_O1} except for orbit $O_3$.} 
\end{figure*}

\begin{figure*}
\includegraphics[scale=0.6]{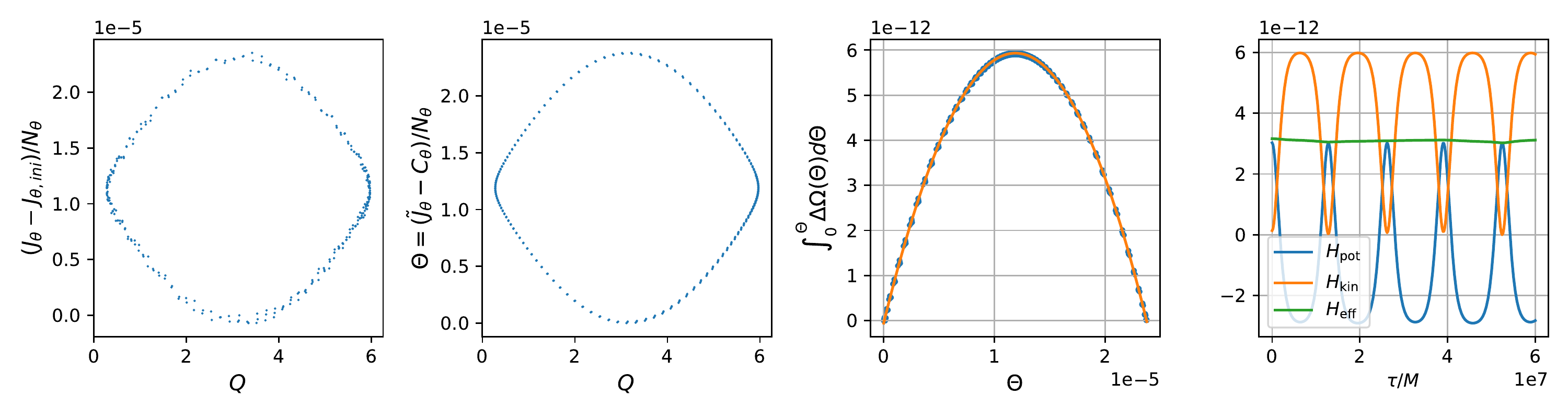}
\caption{\label{fig:phase_O3} Similar to Fig.~\ref{fig:phase_O1} except for orbit $O_3$.} 
\end{figure*}

The behavior of  orbits in the libration regime (orbits that are on the plateau of the rotation curve in Fig.~\ref{fig:Poincare}) is similar to a simple harmonic oscillator, with the resonant angle $Q$ limited to 
a finite range $|\Delta Q| < 2\pi$ (see Figs.~\ref{fig:eff_O3} and \ref{fig:phase_O3} for orbit $O_3$ as an example).
The orbit $O_3$ lies on the edge of the plateau of the rotation curve in Fig.~\ref{fig:Poincare}, whose $\Theta-Q$ phase 
diagram is quite similar to that of orbit $O_2$, with $|\Delta Q|$ smaller than but close to $2\pi$. 
As moving towards the center of the plateau from the left edge, we find the orbital phase diagram becomes more confined in both $\Theta$ and $Q$ directions.  For the orbit exactly lying in the center, both $\Theta$ and $Q$ turn out to be constants, and this point is the so-called stable point.
The on-plateau orbits on the right-hand side (of the stable point) are symmetric to their counterparts on the left-hand side, in the sense that their  orbital phase diagrams $Q-\Theta$ are the same.

\begin{figure}
\includegraphics[scale=0.8]{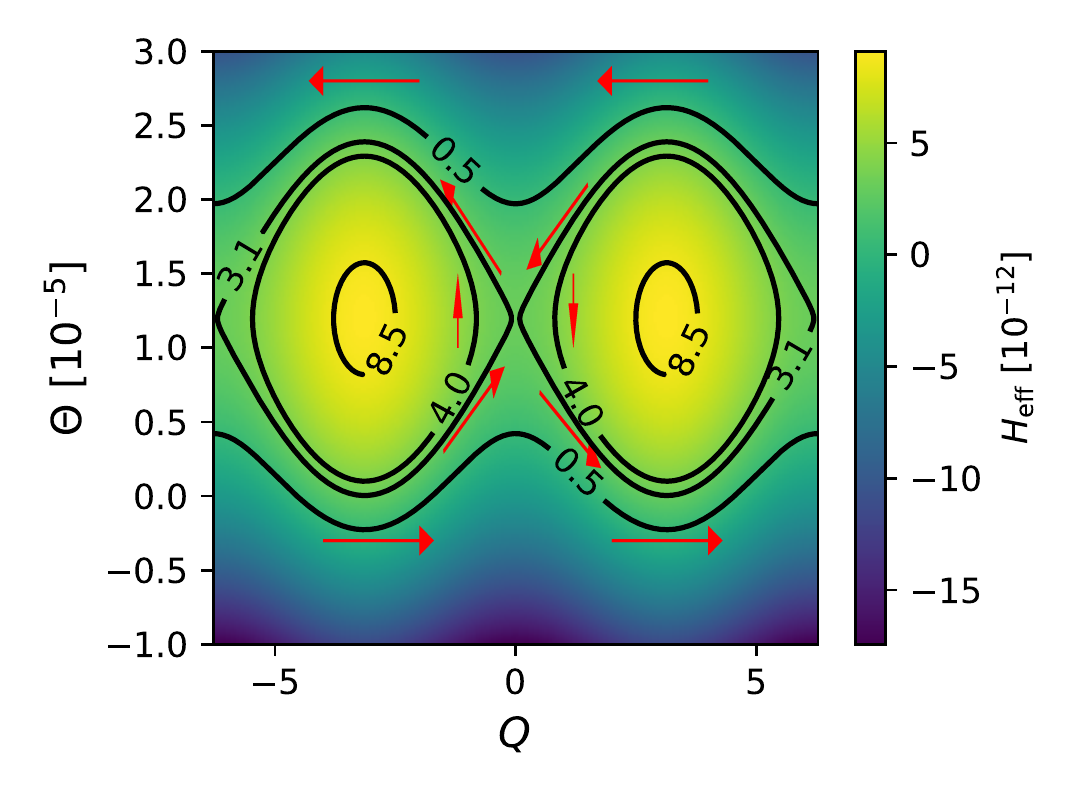}
\caption{\label{fig:contour} Contours of effective Hamiltonian $H_{\rm eff}(Q, \Theta)$ with the red arrows showing their evolution directions.} 
\end{figure}

Orbits belonging to different categories and the symmetry explained above could be summarized with the effective Hamiltonian $\mathcal{H}_{\rm eff} = \alpha_0 \Theta +\beta_0 \Theta^2+H_{\rm res} \cos Q$.
For a given set of parameters $(\alpha_0, \beta_0, H_{\rm res})$, the orbits in the $Q-\Theta$ space are simply contours
of constant effective energy and the orbital behavior (libration, transition, or rotation)
is largely determined by the effective energy. 
As an example, we take $(\alpha_0, \beta_0, H_{\rm res})= (1.00\times10^{-6}, -0.04, 3.03\times10^{-12})$ (the orbit $O_3$ parameters as shown in Figs.~\ref{fig:eff_O3} and \ref{fig:phase_O3}), the critical effective energy turns out to be $\mathcal{H}_{\rm crit} = -\alpha_0^2/(4\beta_0)-H_{\rm res}=2.94\times 10^{-12}$ (c.f. Eq.~\ref{eq:eec}). We show different contour levels in 
Fig.~\ref{fig:contour}. The low-energy contour $\mathcal{H}_{\rm eff} = 0.5\times 10^{-12} < \mathcal{H}_{\rm crit}$ is 
in the rotation regime (e.g., orbit $O_1$). In fact, this orbit consists of two symmetric and disconnected branches,
which correspond to orbit $O_1$ ($\dot Q > 0$) and its symmetric counterpart  ($\dot Q < 0$), respectively. 
The high-energy contour $\mathcal{H}_{\rm eff} = 4\times 10^{-12} > \mathcal{H}_{\rm crit}$ is 
in the libration regime, and this contour corresponds to a pair of orbits that are of the same effective energy while 
opposite rotation directions (e.g., orbit $O_3$ and its symmetric counterpart). The contour of even higher energy $\mathcal{H}_{\rm eff} = 8.5\times 10^{-12}$ corresponds to orbits that are closer to the stable point.
The contour with critical energy partially captures the orbital properties in the transition regime (e.g., $O_2$).
The limitation is that there is no chaos in this one-degree -of-freedom description, because the number of degrees of freedom is equal to the 
number of conserved quantity (the effective energy). In reality, we find the layers of chaos (as previous observed in various spacetimes \cite{Deich:2022vna,Bronicki:2022eqa,Destounis:2020kss,
Lukes-Gerakopoulos:2010ipp}) lying between the rotational and libration orbits, thanks to the contribution from non-resonant terms.

\section{Impact for EMRI evolution}\label{sec:impact}

\begin{figure*}
\includegraphics[width=0.3\textwidth]{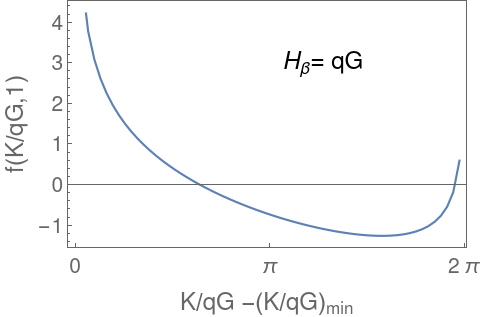}~~~~~~
\includegraphics[width=0.31\textwidth]{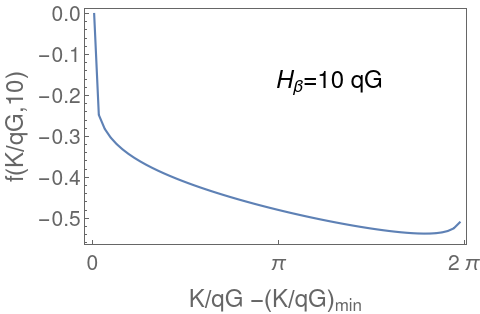}~~~~~~
\includegraphics[width=0.32\textwidth]{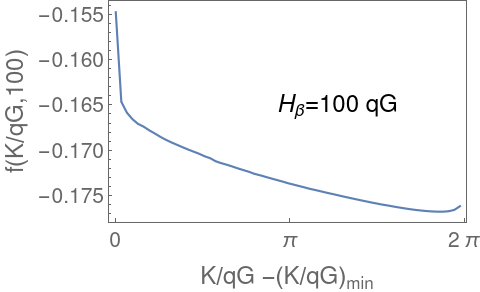}
\caption{\label{fig:plotf} These plots show the function $f(K/qG,H_\beta/qG)$ that appears in $\Delta \tilde{ \mathcal{J}}_\alpha$ in Eq.~\eqref{eq:jturn} for $H_\beta$ fixed to $qG, 10 qG$ and $100 qG$, respectively. For a given value of $H_\beta/qG$, there is a minimum values of $K/qG$ for which $f$ is well-defined, denoted as $(K/qG)_{\rm min}$. The function is $2\pi$-periodic in $K/qG$, so we only plot it for $K/qG-(K/qG)_{\rm} \in \left[ 0, 2\pi \right]$. The shape of the function remains relatively consistent across different values of $H_\beta/qG$, yet the actual values and the range it covers decrease as $H_\beta/qG$ increases.} 
\end{figure*}

\begin{figure*}
\includegraphics[width=0.3\textwidth]{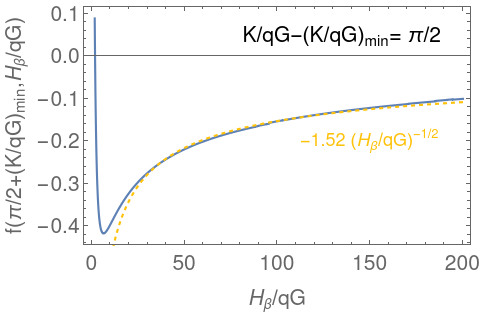}
~~~~~~
\includegraphics[width=0.3\textwidth]{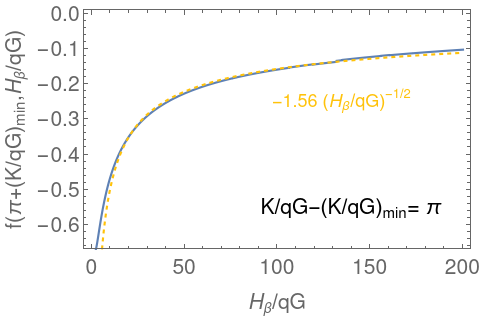}
~~~~~~
\includegraphics[width=0.3\textwidth]{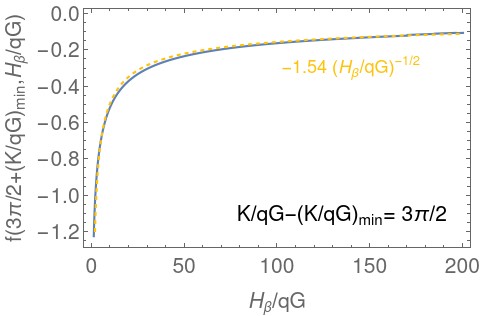}
\caption{\label{fig:plotf-2} Similar to Fig.~\ref{fig:plotf}, these plots depict the function $f(K/qG,H_\beta/qG)$, but now for $K/qG$ fixed. The yellow, dashed lines show numerical fits in the adiabatic regime; for the fits we only used $H_\beta/qG  \in (40,200)$ for the first two plots, while we used the entire plot range for the right plot. } 
\end{figure*}

\begin{figure}
    \centering
    \includegraphics[width=0.4\textwidth]{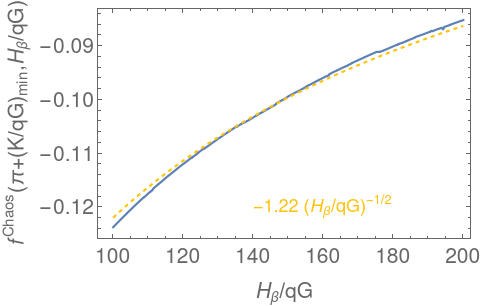}
    \caption{This plot shows the integral in Eq.~(\ref{eq:jturnchaos}) with $Q^c$ set such that $\Delta \Theta_{\rm chaos}$ is $\sqrt{0.1}$ times the size of the critical curve in the $\Theta$ direction, which is $|\alpha_0/\beta_0|$. We have set $K/qG-(K/qG)_{\rm min}=\pi$.}
    \label{fig:jump-chaos}
\end{figure}

The long-term impact of resonance relies on the jump of action variables $\Delta \mathcal{J}_\alpha$, as discussed in \cite{Flanagan:2010cd,Bonga:2019ycj}. First, let us neglect the effect of chaotic transitional orbits, so that Eq.~\eqref{eq:k} can be used to compute the shift of the action variables across a resonance. Introducing $H_\beta=\beta_0 H_{\rm res}$ and $G = \sum_\alpha a_\alpha G_\alpha$, we can rewrite Eq.~\eqref{eq:k} in a more convenient form 
\begin{align}
\frac{1}{2}\left ( \frac{d Q}{d \tau}\right )^2+ H_{\beta} \cos Q +q \, G Q =K\,,
\end{align}
or
\begin{align}
\frac{d Q}{d \tau}= \pm \sqrt{2}\sqrt{K-H_{\beta}\cos Q -q G Q}
\end{align}
where the sign depends on the branch the orbit belongs to (e.g., see Fig.~\ref{fig:contour}). This equation determines the $\tau$ dependence of the resonant angle $Q$. The additional kicks on $\tilde{\mathcal{J}}_\alpha$, according to Eq.~\eqref{eq:sim}, are given by
\begin{align}\label{eq:Deltaja}
\Delta \tilde{\mathcal{J}}_\alpha = -N_\alpha H_{\rm res}\int^{\infty}_{-\infty} \sin Q(\tau) d\tau
\end{align}
where $\tau=-\infty$ is assumed to be sufficiently before the resonance and $\tau=\infty$ is assumed to be sufficiently after the resonance. The above equation immediately suggests that the discontinuities of $\tilde{\mathcal{J}}_\alpha$ across the resonance are proportional to $N_\alpha$, a phenomenon also found in \cite{Silva:2022blb,Gupta:2022jdt}. This point is important for incorporating resonance effects into EMRI waveforms as it demonstrates nicely that only one free parameter is needed to search for these resonances instead of multiple free parameters. 

In the non-adiabatic limit where $H_{\beta} \ll q G$, the solution of $Q(\tau)$ can be approximated as $Q\approx Q_0+q G \tau^2/2$, for which $\tau=0$ is set at the point that $dQ/d\tau =0$. As a result, we have
\begin{align}\label{eq:nona}
\Delta \tilde{\mathcal{J}}_\alpha = -N_\alpha H_{\rm res} \sqrt{\frac{\pi}{q G}} \left( \cos Q_0 + \sin Q_0 \right)
\end{align}
which is consistent with \cite{Bonga:2019ycj,Flanagan:2010cd}. Notice that in this limit the trajectory may temporarily cross the resonant islands in the phase space  because the resonance capture condition derived for the parametric change of the effective Hamiltonian (see \cite{murray1999solar}) requires adiabatic evolution. 

In general, we can rewrite Eq.~\eqref{eq:Deltaja} as
\begin{align}\label{eq:jturn}
\Delta \tilde{\mathcal{J}}_\alpha & = -N_\alpha H_{\rm res}\int^{\infty}_{-\infty} \sin Q(\tau) \frac{dQ}{dQ/d\tau} \nonumber \\
& =  -\sqrt{2} N_\alpha H_{\rm res}\int^{Q^*}_{-\infty} d Q\frac{\sin Q }{\sqrt{K-H_{\beta}\cos Q -q G Q}} \nonumber \\
& = -\sqrt{2} N_\alpha \frac{H_{\rm res}}{\sqrt{q G}} f(K/qG, H_\beta/qG)\,,
\end{align}
where at $Q=Q^*$ we assume that $dQ/d\tau=0$, so that $K=H_\beta \cos Q^* +q G Q^*$. This function is shown in Fig.~\ref{fig:plotf} with $H_\beta/qG$ fixed to three different values, and in Fig.~\ref{fig:plotf-2} $K/qG$ fixed to three different values. In the adiabatic limit $H_\beta \gg q G$, we can approximate $f \sim \mathcal{O}(1)(H_\beta/q G)^{-1/2}$, so that the overall $\Delta \tilde{\mathcal{J}}_\alpha$ is proportional to $(H_\beta)^{1/2}$.

With the chaotic transitional orbits considered, the orbit may jump from one branch to the other branch before reaching the turning point ($d Q/d\tau =0$), as shown in Fig.~\ref{fig:phase_O2}. In this case, the trajectory makes the transition between branches along with a jump in $\Theta$: $\Delta \Theta_{\rm chaos} = \Delta \tilde{\mathcal{J}}_{\alpha, {\rm chaos}}/N_\alpha$. The corresponding turning $Q^*$ in Eq.~\eqref{eq:jturn} should be replaced by
\begin{align}
K= H_{\beta} \cos Q^c + q G Q^c +\frac{(\beta_0 \Delta \Theta_{\rm chaos})^2}{8}
\end{align}
so that the accumulated shift in $\tilde{\mathcal{J}}_\alpha$ before and after the transition is
\begin{align}\label{eq:jturnchaos}
\Delta \tilde{\mathcal{J}}_\alpha &=
 -\sqrt{2} N_\alpha H_{\rm res}\int^{Q^c}_{-\infty} d Q\frac{\sin Q }{\sqrt{K-H_{\beta}\cos Q -q G Q}} +\Delta \tilde{\mathcal{J}}_{\alpha, {\rm chaos}}\nonumber \\
& =\Delta \tilde{\mathcal{J}}^c_\alpha+ \Delta \tilde{\mathcal{J}}_{\alpha, {\rm chaos}}\,.
\end{align}

Because the non-resonant terms and resonant terms are generated by the same $h$, we expect that the width of the chaotic zone is a fraction of the size of the resonant island, which is proportional to $\sqrt{H_{\rm res}}$. As a result, we expect that $\Delta \tilde{\mathcal{J}}_{\alpha, {\rm chaos}} \propto \sqrt{H_{\rm res}}$. The numerical value of $\Delta \Theta_{\rm chaos}$ in Fig. ~\ref{fig:phase_O2} is approximately $10\%$ of the size of the critical resonant island in the $\Theta$ direction (also approximately true for the jump in the $1/2$ resonance case shown in Fig.~\ref{fig:app3}), which is given by $|\alpha_0/\beta_0|$.
This number may not be universal for  resonances, because we find that even for the same system, the transitional with longer evolution tends can give rise to larger jumps. Nevertheless it can serve as a rough estimate for the impact of chaos. The size of $\Delta \Theta_{\rm chaos}$ determines the upper limit $Q^c$ in the evaluation for $\Delta \mathcal{J}^c_\alpha$. For numerical illustration purpose, in Fig.~\ref{fig:jump-chaos} we have increased $\Delta \Theta_{\rm chaos}$ such that it is $\sqrt{0.1}$ times $|\alpha_0/\beta_0|$.  We find that $\Delta \mathcal{J}^c_\alpha$ follows a similar scaling law as $\Delta \tilde{\mathcal{J}}_\alpha$ in Eq.~(\ref{eq:jturn}). As we decrease the magnitude of $\Delta \Theta_{\rm chaos}$, we expect that the coefficient of the scaling law will be closer to the ones shown in Fig.~\ref{fig:plotf-2}.

Overall, the long-term phase shift due to resonance should be
\begin{align}
\Delta \Psi \sim \frac{\Delta \mathcal{J}}{\mathcal{J}} \frac{1}{q}
\end{align}
where the $1/q$ factor comes from the $1/q$ scaling of the radiation reaction timescale. Accordingly we conclude that in the non-adiabatic regime, the transient resonance crossing gives rise to
\begin{align}
\Delta \Psi \sim \frac{\mathcal{O}(H_{\rm res})}{q^{3/2}} \,.
\end{align}

In the adiabatic regime, the chaotic transitional orbits generally gives rise to a phase shift
\begin{align}
\Delta \Psi_{\rm chaos} \sim \mathcal{O}(0.1)\frac{H_{\rm res}^{1/2}}{q}
\end{align}
and the total induced-phase shift including the additional change of $\mathcal{J}_\alpha$ before and after reaching the transitional orbits should approximately scale as
\begin{align}
\Delta \Psi  \sim \frac{\mathcal{O}(H_{\rm res}^{1/2})}{q}\,.
\end{align}

\section{Discussion and Conclusion}\label{sec:conclusion}


An EMRI system may be influenced by environmental gravitational perturbations, which can come from nearby astrophysical objects, dense dark matter distributions and/or exotic compact objects. Because of the large number of orbital cycles in band ($10^4-10^5$) for space-borne gravitational wave detectors, EMRIs will provide unprecedented opportunities to probe these environmental forces. 
To enable waveform building for upcoming gravitational wave experiments, it is necessary to comprehend the EMRI evolution and waveform on perturbed Kerr backgrounds.

The work presented here represents one part of this program, in which resonance effects on the EMRI dynamics and its impact on the waveform is analyzed by introducing a resonance effective Hamiltonian. With this formalism, one can make connection with extensive literature on mean-motion resonance in planetary systems. Future studies along this direction will be interesting, especially in terms of understanding the transitional chaotic orbits from the Hamiltonian point of view. On the other hand, for realistic EMRI evolution, the system likely falls into the non-adiabatic regime for resonance crossing (except for relatively large $h$), as discussed in Sec.~\ref{sec:impact}. Since the numerical investigation in Sec.~\ref{sec:nu} have not included gravitational radiation reaction, it will also be interesting to perform the numerical studies with radiation reaction included to verify the corresponding statements in Sec.~\ref{sec:rcross} and Sec.~\ref{sec:impact}. 

The other part of the program, i.e. understanding the EMRI evolution in the non-resonance regime of a perturbed Kerr background, will be discussed in a separate work. By combining the results in these studies, one should be able to build an EMRI waveform model (on top of the EMRI waveform model on a Kerr background) for a generic Kerr perturbation $h$. 

This method is not only important for future space-borne gravitational wave detection, but likely also useful for constructing waveforms to test modified gravity theories using comparable mass-ratio binaries in the band of ground-based gravitational wave detectors. Indeed, for waveforms consistent with General Relativity, there is literature \cite{LeTiec:2011bk,LeTiec:2011dp,LeTiec:2013uey,vandeMeent:2016hel,vandeMeent:2020xgc,Rifat:2019ltp,Feng:2021sax} pointing out that the EMRI-inspired waveform, with an appropriate rescaling  using the mass-ratio parameter, agrees surprisingly well with the waveform obtained from numerical relativity for comparable mass-ratio systems. As the EMRI method does not require the post-Newtonian or post-Minkowskian expansion, it will serve as a promising route for generating high-precision waveforms for comparable mass-ratio binaries. It is also natural to expect that a modified gravity waveform produced with the EMRI approach may also inherit similar advantages. This application may be potentially important as presently we have limited waveform models going beyond the post-Newtonian approximation, partially because of the well-posedness issue in many modified theories of gravity.

In this work, we have assumed axisymmetric for the metric perturbation $h$. If $h$ is non-axisymmetric, the analysis in Sec.~\ref{sec:Heff} and Sec.~\ref{sec:impact} should still apply except that we should include more general resonances with nonzero $N_\phi$, similar to the tidal resonance considered in \cite{Bonga:2019ycj}. Because of the expanded space of relevant resonances, it becomes more likely that during the evolution of an EMRI system, multiple resonances will be important. If the EMRI crosses the resonances in a non-adiabatic manner, from Eq.~(\ref{eq:nona}) we see that there is a phase $Q_0$ that is related to the initial phase of the orbit. It will be interesting to investigate whether $Q_0$ of later resonances can still be determined accurately, with the earlier resonance jumps influencing the orbit on the radiation reaction timescale.

\acknowledgments

We thank Mohammed Khalil for helpful discussions.
Z. P. and H. Y. are supported by the Natural Sciences and
Engineering Research Council of Canada and in part by
Perimeter Institute for Theoretical Physics.
L. B. is grateful for the hospitality of Perimeter Institute where part of this work was carried out. Research at Perimeter Institute is supported in part by the Government of Canada through the Department of Innovation, Science and Economic Development Canada and by the Province of Ontario through the Ministry of Colleges and Universities. This research was also supported in part by the Simons Foundation through the Simons Foundation Emmy Noether Fellows Program at Perimeter Institute.
L. B. also acknowledges financial support by the ANR PRoGRAM project, grant ANR-21-CE31-0003-001, and from the EU Horizon 2020 Research and Innovation Programme under the Marie Sklodowska-Curie Grant Agreement no. 101007855.

\appendix

\section{$1/2$ Resonance}\label{sec:app1}
In the main text, we have been focusing on orbits that are close to the $2/3$ resonance, i.e., $\nu\approx 2/3$.
In this appendix, we examine the  orbits that are close to the $1/2$ resonance, which share some common features with those $2/3$-resonance orbits, but also yield some differences.

\begin{figure*}
\includegraphics[scale=0.8]{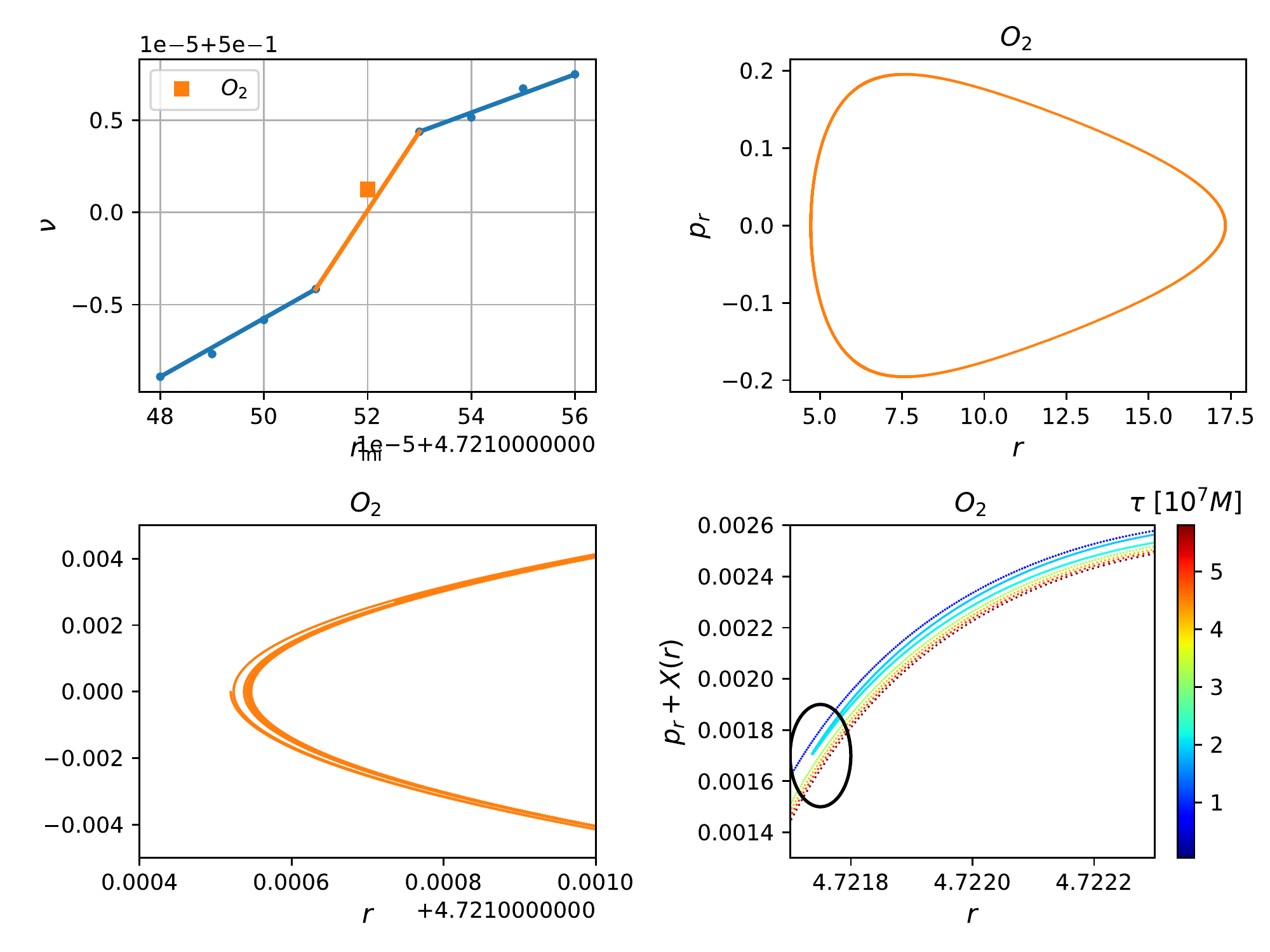}
\caption{\label{fig:app1}
Near-resonance orbits in the perturbed Kerr spacetime with parameter $(a, \epsilon)=(0.2, 0.002)$.
Upper left panel: the rotation numbers of near-resonance orbits, with $(E, L)=(0.96, 3.5)$ and initial conditions $(\theta, p_r)_{\rm ini}=(\pi/2, 0)$, where the dots are the numerical results and the straight lines of two different colors
denote two different kinds of orbits: regular orbits in blue and chaotic transitional orbits in orange, and the square dot are the representative orbit  $O_{2}$  with initial radius $r_{\rm ini}=4.72152$. Upper right panel: Poincar\'e map $(r, p_r)|_{\theta=\pi/2}$ of orbits $O_{2}$, respectively. 
Lower left panel: zoomed-in Poincar\'e map of  orbit $O_2$, where we see two branches.
Lower right panel: further zoomed-in Poincar\'e map of  orbit $O_2$, 
and $X(r)=-3\times(r-4.7214)$, where a turn-back shows up around $r=4.72$ at $\tau\approx2\times10^7 M$.} 
\end{figure*}

\begin{figure*}
\includegraphics[scale=0.8]{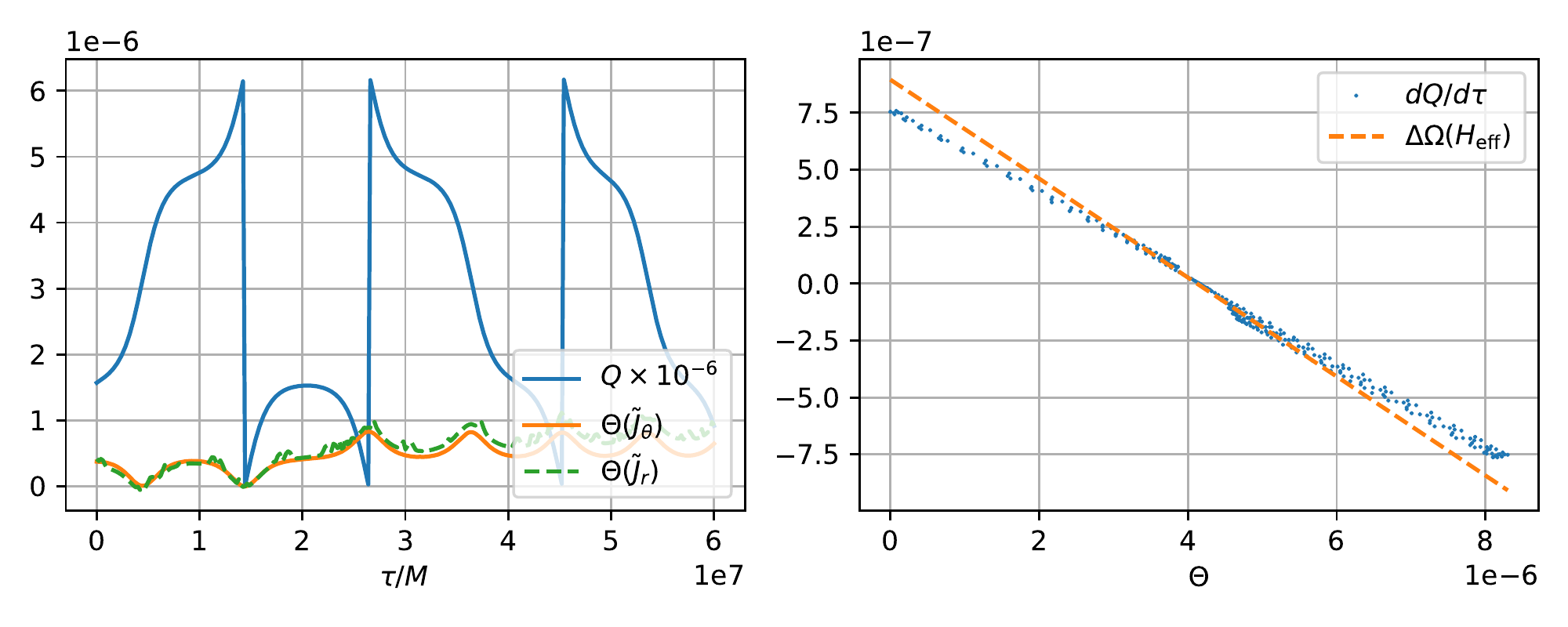}
\caption{\label{fig:app2} Similar to Fig.~\ref{fig:eff_O2}, except for the transitional orbit $O_2$ that is close to the $1/2$ resonance.} 
\end{figure*}

\begin{figure*}
\includegraphics[scale=0.6]{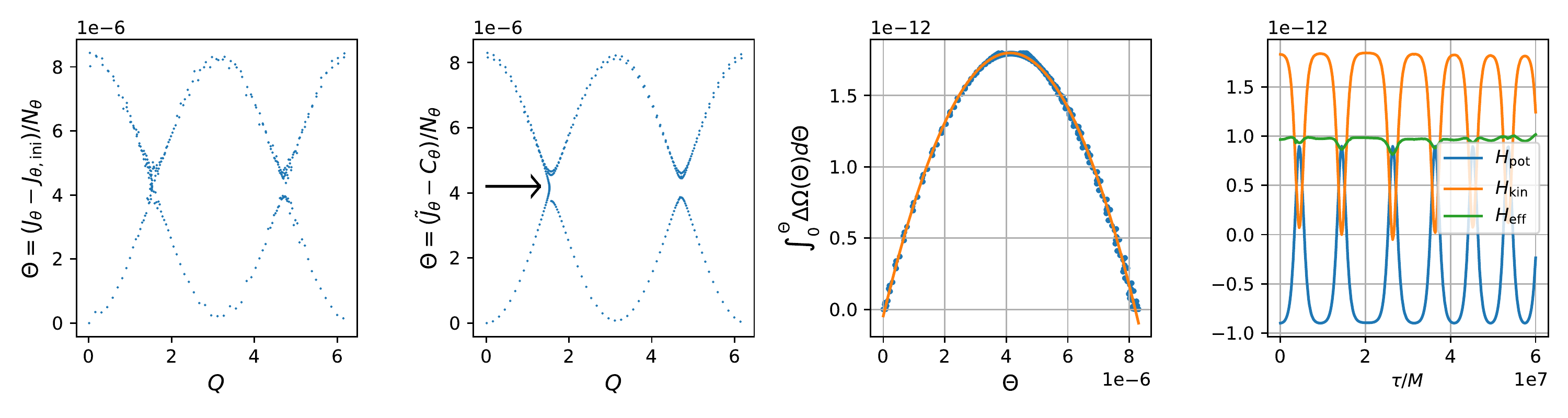}
\caption{\label{fig:app3} Similar to Fig.~\ref{fig:phase_O2}, except for the transitional orbit $O_2$ that is close to the $1/2$ resonance.} 
\end{figure*}

Similar to the $2/3$-resonance, we consider the perturbed Kerr spacetime with parameters $(a, \zeta)=(0.2, 0.002)$
and orbits with energy and angular momentum $(H, E, L)=(-0.5, 0.96, 3.5)$ and initial conditions $(\theta, p_r)_{\rm ini}=(\pi/2, 0)$, $r_{\rm ini}\approx 4.72 M$.  In the top left panel of Fig.~\ref{fig:app1}, we show the rotation numbers of the orbits (labelled by $r_{\rm ini}$) that are close to the $1/2$ resonance. Different from the $2/3$-resonance case, there is no plateau appearing in the rotation curve, i.e., there is no orbit in the libration regime. The remaining near-resonance orbits can be classified into two categories:  orbits in the rotation regime (blue line) and chaotic transitional orbits (orange line).
In the remaining three panels, we show the Poincar\'e map $(r, p_r)_{\theta=\pi/2}$ of orbit $O_2$ and two zoomed-in versions.
In the lower two panels, we see that the Poincar\'e maps consists of two branches, a $Q$ increasing branch (the ``counter-clockwise" branch in Fig.~\ref{fig:app1}) and a $Q$ decreasing branch (the ``clockwise" branch in Fig.~\ref{fig:app1}) with the transition occurring around $\tau=2\times10^7 M$. Similar to the $2/3$-resonance case, this transition is also marked by a jump in momentum $\Theta$ (see Figs.~\ref{fig:app2} and \ref{fig:app3}).

The major difference between the $2/3$-resonance orbits considered in the main text and the $1/2$-resonance orbits considered here is most easily understood from their Poincar\'e maps. The $2/3$-resonance orbits cross one of the islands and the island center, which is a stable point, while the $1/2$ orbits avoid the islands by crossing the adjunction point between two islands, which is an unstable point.
As a result, the $1/2$-resonance orbits pass the resonance without being trapped on the resonance, i.e., no orbit has $\nu=1/2$.

If we consider more general initial conditions for distinct sets of orbits, such as $(p_r)_{\rm ini} \neq 0$, we generally observe that the orbits will cross one of the islands, but without crossing the island center (the stable point) or the adjunction point between two islands (the unstable point).


\section{$2/3$ Resonance with stronger perturbation}\label{sec:appx}

\begin{figure*}
\includegraphics[scale=0.8]{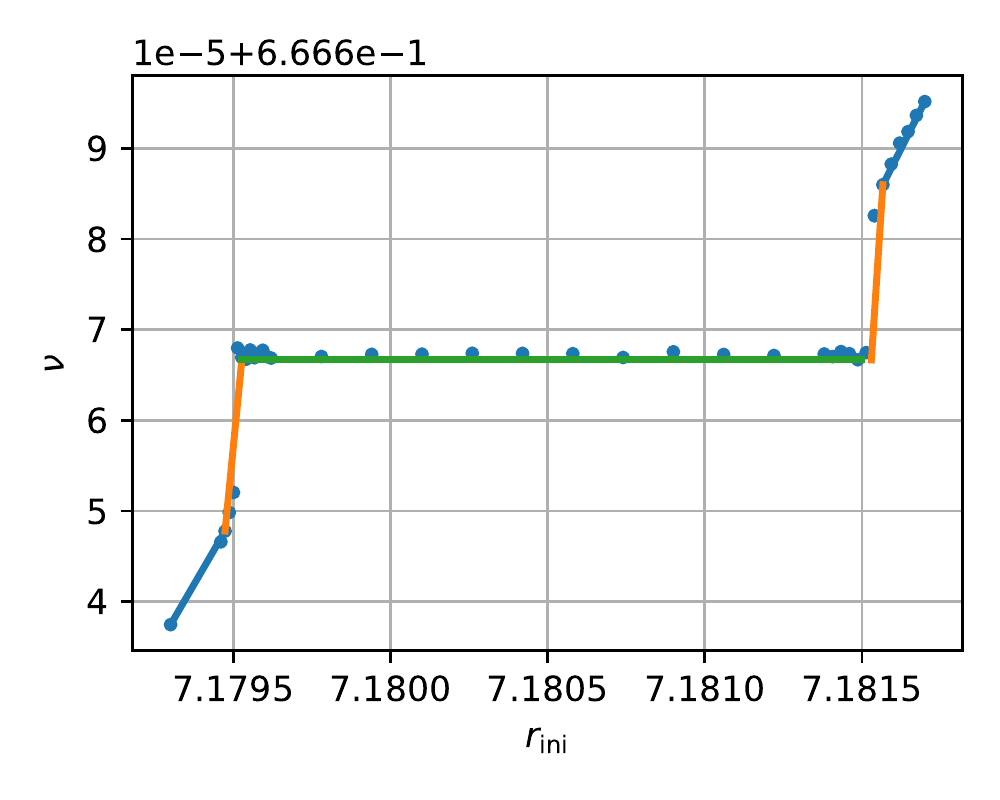}
\caption{\label{fig:appx} Similar to the first panel in Fig.~\ref{fig:Poincare}, except with $\zeta=0.02$.} 
\end{figure*}

In order to verify the $\sqrt{\epsilon}$ dependence of the size of the resonant regimes and the chaotic regime in the phase space, as suggested by the effective Hamiltonian formalism, we increase $\zeta$ from $0.002$ in  Fig.~\ref{fig:Poincare} to $\zeta=0.02$ in Fig.~\ref{fig:appx}. It is  evident that the width of the plateau (proportional to the size of the resonant island) and the height of the segment near the plateau (proportional to the size of the chaotic regime) are both approximately increased by $\sqrt{10}$.

\section{Mapping to Action-Angle Variables $(q, J)$}\label{sec:app2}
\label{app:to-action-angle-variables}

The necessary steps for mapping to action-angle variables $(q, J)$ in Kerr spacetime has been investigated in detail in  Refs.~\cite{Schmidt2002,Hinderer2008}, and we closely follow their discussions in this work. 
With this mapping in hand, in principle 
one can evolve the equation of motion in terms of action-angle variables [Eq.~(\ref{eq:eom})].
But as we will see later, the mapping is mostly done numerically, which is slow and not sufficiently accurate 
in the long-term evolution for pinning down the small near-resonance features.
Therefore, we choose to perform the evolution in physical coordinates $(x, p)$, and then mapped them to action-angle variables $(q, J)$.

In the Kerr spacetime, the energy and the $z$-component of angular momentum 
are 
\be 
E = -p_t\ , 
\ee 
and 
\be 
L = p_\phi\ .
\ee 
The Carter constant is \cite{Carter1968}
\be \label{eq:Carter}
C = p_\theta^2 + a^2\cos^2\theta(\mu^2-p_t^2)+\cot^2\theta p_\phi^2\ ,
\ee 
and the Hamiltonian is 
\be 
\begin{aligned}
   H = \frac{1}{2}g^{\mu\nu}p_\mu p_\nu = &\frac{\Delta}{2\Sigma}p_r^2 + \frac{1}{2\Sigma} p_\theta^2
+\frac{(p_\phi+a\sin^2\theta p_t)^2}{2\Sigma\sin^2\theta} \\
&-\frac{[(r^2+a^2)p_t+ap_\phi]^2}{2\Sigma\Delta}\ .
\end{aligned}
\ee 
For mapping the physical coordinates to the action-angle variables, 
we need to invert the four equations above and write the momentum $p_\mu$ in terms of the integrals of motion as 
\be \label{eq:momenta}
\begin{aligned}
    p_r &= \pm\frac{\sqrt{V_r(r)}}{\Delta}\ , &p_\theta &= \pm \sqrt{V_\theta(\theta)} \ , \\ 
    p_t &= -E\ , &p_\phi &= L \ ,
\end{aligned}
\ee 
where the two potentials are  
\be 
\begin{aligned}
    V_r(r) &= [(r^2+a^2)E-aL]^2-\Delta[\mu^2r^2+(L-aE)^2+C]\ , \\
    V_\theta(\theta) &= C-\left[(\mu^2-E^2)a^2+\frac{L^2}{\sin^2\theta}\right]\cos^2\theta\ .
\end{aligned}
\ee 
Given the integrals of motion  $P_\alpha=(H, E, L, C)$, one can obtain the pericenter/apocenter separations $r_{-/+}$
as the two largest roots to equation $V_r(r)=0$, and obtain the minimum/maximum polar angles $\theta_{-/+}$
as the roots to equation equation $V_\theta(\theta)=0$ satisfying the constraint $\cos^2\theta \in [0, 1]$.
In terms of commonly used orbital parameters, eccentricity $e$ and semi-latus rectum $p$, $r_\pm=p/(1\mp e)$.

The actions are defined as 
\be 
\begin{aligned}
    J_r &= \frac{1}{2\pi}\oint p_r dr \ ,  \\ 
    J_\theta &=\frac{1}{2\pi}\oint p_\theta d\theta \ ,  \\ 
    J_t &= \frac{1}{2\pi}\int_0^{2\pi} p_t dt = -E \ , \\
    J_\phi &=\frac{1}{2\pi}\oint p_\phi d\phi = L \ ,
\end{aligned}
\ee 
and the numerical integration above is straightforward with  the momenta defined in Eq.~(\ref{eq:momenta}).

To obtain the angle variables, we need the canonical transformation from the physical coordinates $(x^\mu, p_\nu)$
to $(q^\alpha, J_\beta)$ associated with a general solution to the Hamilton-Jacobi equation 
\be 
H\left(x^\mu, \frac{\partial\mathcal{S}}{\partial x^\mu}\right) + \frac{\partial \mathcal{S}}{\partial\tau} = 0\ .
\ee 
Since the Hamiltonian does not involve time $\tau$ explicitly, the Hamilton's principle function $\mathcal{S}$ 
can be written in the form 
\be 
\mathcal{S} = \gamma\tau + \mathcal{W}(x^\mu, P_\alpha)\ ,
\ee 
where $\gamma=\mu^2/2$ and  $\mathcal{W}(x^\mu, P_\alpha)$ is called Hamilton's characteristic function.
As shown by Carter \cite{Carter1968}, 
\be \label{eq:W}
\mathcal{W}(x^\mu, P_\alpha) = - Et +L\phi \pm \mathcal{W}_r(r) \pm \mathcal{W}_\theta(\theta)\ ,
\ee 
where
\be \label{eq:W_rtheta}
\begin{aligned}
\mathcal{W}_r(r) &= \int^r dr \frac{\sqrt{V_r}}{\Delta}\ , \\ 
\mathcal{W}_\theta(\theta) &= \int^\theta d\theta \sqrt{V_\theta}\ .   
\end{aligned}
\ee 
Taking $\mathcal{W}(x^\mu, J_\beta)$ as a generating function that generates a canonical transform from 
the physical coordinates $(x^\mu, p_\nu)$ to $(q^\alpha, J_\beta)$ (Chapter 9 of \cite{Goldstein2002}), we have 
\be \label{eq:pq}
\begin{aligned}
    p_\nu &=\frac{\partial \mathcal{W}}{\partial x^\nu}(x^\mu, J_\beta)\ , \\
    q^\alpha &= \frac{\partial \mathcal{W}}{\partial J_\alpha}(x^\mu, J_\beta)
    = \frac{\partial \mathcal{W}}{\partial P_\beta}\frac{\partial P_\beta}{\partial J_\alpha}\ .
\end{aligned}
\ee 
There is some sign uncertainty in the definition of $\mathcal{W}(x^\mu, P_\alpha)$ [Eq.~(\ref{eq:W})], 
which is fixed by the first line in Eq.~(\ref{eq:pq}), i.e., we fix the sign uncertainty with 
\be 
\mathcal{W} = - Et +L\phi +{\rm sgn}(p_r) \mathcal{W}_r(r) +{\rm sgn} (p_\theta) \mathcal{W}_\theta(\theta)\ ,
\ee 
in calculating $q^\alpha $ from the second line in Eq.~(\ref{eq:pq}). Then the calculation of $q^\alpha$ consists of three major components: $\partial P_\beta/\partial J_\alpha$ and $\partial\mathcal{W}_{r,\theta}/\partial P_\alpha$.

\subsection{$\partial J_\beta/\partial P_\alpha$ and $\partial P_\beta/\partial J_\alpha$ }
The analytic expressions of the partial derivatives $\partial J_\beta/\partial P_\alpha$ had been detailed in~\cite{Schmidt2002,Hinderer2008}, 
and we repeat the non-trivial components for convenience here:
\be 
\begin{aligned}
    \frac{\partial J_r}{\partial H} &= \frac{Y(r_-,r_+)}{\pi}\ , \\ 
    \frac{\partial J_r}{\partial E} &= \frac{W(r_-,r_+)}{\pi}\ , \\
    \frac{\partial J_r}{\partial L} &= -\frac{Z(r_-,r_+)}{\pi}\ , \\ 
    \frac{\partial J_r}{\partial C} &= -\frac{X(r_-,r_+)}{2\pi}\ ,  
\end{aligned}
\ee 
and 
\be 
\begin{aligned}
     \frac{\partial J_\theta}{\partial H} &= -\frac{2\sqrt{z_+}a^2}{\pi\beta}[K(k)-E(k)]\ , \\ 
    \frac{\partial J_\theta}{\partial E} &= -\frac{2\sqrt{z_+}Ea^2}{\pi\beta}[K(k)-E(k)]\ , \\ 
    \frac{\partial J_\theta}{\partial L} &= -\frac{2L}{\pi\beta\sqrt{z_+}}[K(k)-\Pi(z_-,k)]\ , \\
    \frac{\partial J_\theta}{\partial C} &= -\frac{1}{\pi\beta\sqrt{z_+}}K(k)\ , \\ 
\end{aligned}
\ee 
where $W, X, Y, Z$ are defined as \cite{Schmidt2002}
\be 
\begin{aligned}
    W(r_-,r_+) &= \int_{r_-}^{r_+} \frac{r^2E(r^2+a^2)-2Mra(L-aE)}{\Delta\sqrt{V_r}} dr\ ,\\
    X(r_-,r_+) &= \int_{r_-}^{r_+} \frac{1}{\sqrt{V_r}} dr\ ,\\
    Y(r_-,r_+) &= \int_{r_-}^{r_+} \frac{r^2}{\sqrt{V_r}} dr\ ,\\
    Z(r_-,r_+) &= \int_{r_-}^{r_+} \frac{r[Lr-2M(L-aE)]}{\Delta\sqrt{V_r}} dr\ .
\end{aligned}
\ee 
Here $K(k)$ is the complete elliptic integral of the first kind, $E(k)$ is the complete elliptic integral of the second kind ,
and $\Pi(n,k)$ is the Legendre elliptic integral of the third kind:
\be 
\begin{aligned}
    K(k) &= \int_0^{\pi/2} \frac{d\psi}{\sqrt{1-k^2\sin^2\psi}} \ , \\
    E(k) &= \int_0^{\pi/2} d\psi \sqrt{1-k^2\sin^2\psi} \ , \\
    \Pi(n,k)&=  \int_0^{\pi/2}\frac{d\psi}{(1-n\sin^2\psi)\sqrt{1-k^2\sin^2\psi}} d\psi \ , 
\end{aligned}
\ee 
where $\beta=a\sqrt{\mu^2-E^2}$, $k=\sqrt{z_-/z_+}$, with $z=\cos^2\theta$ and $z_{\pm}$ are the two roots to $V_\theta(z)=0$
with $0< z_-< 1 <z_+$.

With $\partial J_\beta/\partial P_\alpha$, its inverse matrix $\partial P_\beta/\partial J_\alpha$ can be obtained 
either numerically or analytically, where the analytic formula of the angular frequencies 
\be
\Omega^\alpha := \frac{\partial H}{\partial J_\alpha}
\ee 
have been derived in \cite{Hinderer2008} and are
\be 
\begin{aligned}
    \Omega^t &= \frac{K(k)W(r_-,r_+)+a^2z_+E[K(k)-E(k)]X(r_-,r_+)}{K(k)Y(r_-,r_+)+a^2z_+[K(k)-E(k)]X(r_-,r_+)}\ , \\  
    \Omega^r &= \frac{\pi K(k)}{K(k)Y(r_-,r_+)+a^2z_+[K(k)-E(k)]X(r_-,r_+)}\ ,  \\
    \Omega^\theta &=  \frac{\pi\beta\sqrt{z_+}X(r_-,r_+)/2}{K(k)Y(r_-,r_+)+a^2z_+[K(k)-E(k)]X(r_-,r_+)}\ , \\  
    \Omega^\phi &= \frac{K(k)Z(r_-,r_+)+L[\Pi(z_-,k)-K(k)]X(r_-,r_+)}{K(k)Y(r_-,r_+)+a^2z_+[K(k)-E(k)]X(r_-,r_+)}\ . 
\end{aligned}
\ee 

\subsection{$\partial\mathcal{W}_{r}/\partial P_\alpha$ and $\partial\mathcal{W}_{\theta}/\partial P_\alpha$}
Comparing the definition of  $J_r$ with that of $\mathcal{W}_{r}$, it is evident that 
$\partial\mathcal{W}_{r}/\partial P_\alpha$ is similar to $\partial J_{r}/\partial P_\alpha$:
\be 
\begin{aligned}
    \frac{\partial \mathcal{W}_r}{\partial H} &= \frac{Y(r_-,r)}{\pi}\ , \\ 
    \frac{\partial \mathcal{W}_r}{\partial E} &= \frac{W(r_-,r)}{\pi}\ , \\
    \frac{\partial \mathcal{W}_r}{\partial L} &= -\frac{Z(r_-,r)}{\pi}\ , \\ 
    \frac{\partial \mathcal{W}_r}{\partial C} &= -\frac{X(r_-,r)}{2\pi}\ .
\end{aligned}
\ee 

In a similar way, comparing the definition of  $J_\theta$ with that of $\mathcal{W}_{\theta}$, we have 
\be 
\begin{aligned}
    \frac{\partial \mathcal{W}_\theta}{\partial H} &= -\frac{\sqrt{z_+}a^2}{\pi\beta}[K(\phi, k)-E(\phi, k)]\ , \\ 
    \frac{\partial \mathcal{W}_\theta}{\partial E} &= -\frac{\sqrt{z_+}Ea^2}{\pi\beta}[K(\phi, k)-E(\phi, k)]\ , \\ 
    \frac{\partial \mathcal{W}_\theta}{\partial L} &= -\frac{L}{\pi\beta\sqrt{z_+}}[K(\phi, k)-\Pi(\phi,z_-,k)]\ , \\
    \frac{\partial \mathcal{W}_\theta}{\partial C} &= -\frac{1}{2\pi\beta\sqrt{z_+}}K(\phi, k)\ , \\ 
\end{aligned}
\ee 
where
\be 
\begin{aligned}
    K(\phi, k) &= \int_\phi^{\pi/2} \frac{d\psi}{\sqrt{1-k^2\sin^2\psi}} \ , \\ 
    E(\phi, k) &= \int_\phi^{\pi/2} d\psi \sqrt{1-k^2\sin^2\psi}\ ,  \\ 
    \Pi(\phi, n,k)&=  \int_\phi^{\pi/2}\frac{d\psi}{(1-n\sin^2\psi)\sqrt{1-k^2\sin^2\psi}}  \ , 
\end{aligned}  
\ee 
and $\phi = \phi(\theta)= \arcsin(\cos(\theta)/\sqrt{z_-})$, i.e., $\phi(\theta=\pi/2)=0$ and $\phi(\theta=\theta_-)=\pi/2$.

Note that  the lower integration limits in $\mathcal{W}_{r,\theta}$ [Eq.~(\ref{eq:W_rtheta})]
are undefined, as a result, the mapping $q^\alpha(x^\mu, p_\nu)$ can only be determined up to some constant for a given $P_\beta$ or equivalently $J_\alpha$.
In this work, we have fixed the lower integration limits to $r_-$ and $\theta_-$, respectively, i.e., 
we have fixed the mapping freedom by choosing $q^r(r=r_-)=0$ and $q^\theta(\theta=\theta_-)=0$.
With all the equations summarized above, we can numerically obtain the mapping $q^\alpha(x^\mu, p_\nu)$ and $J_\alpha(x^\mu, p_\nu)$.

\section{ Near Identity Transformation $(q, J)\rightarrow (\tilde q, \tilde J)$}\label{sec:app3}

\begin{figure*}
\includegraphics[scale=0.52]{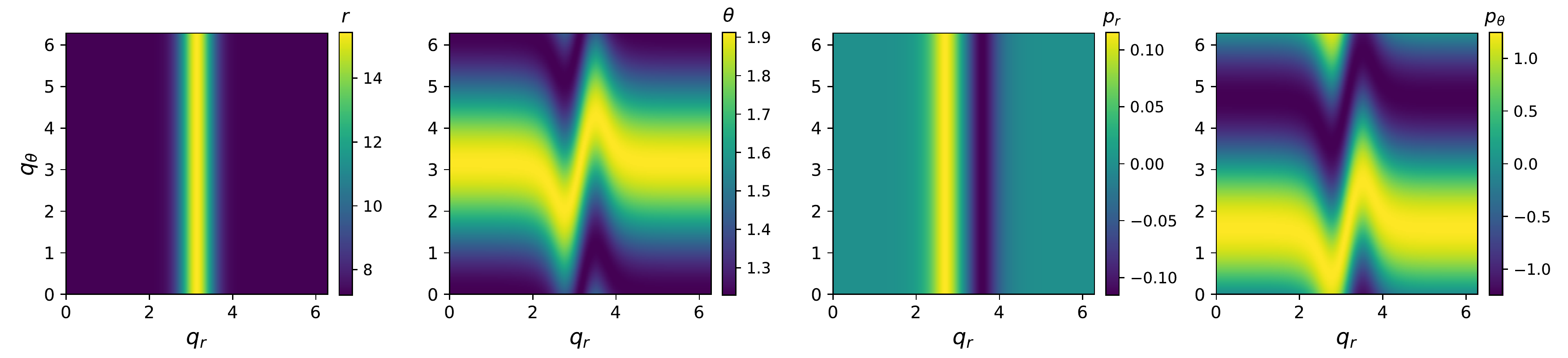}
\caption{\label{fig:app4} $(r, \theta, p_r, p_\theta)$ as functions of $q^{r}$ and $q^\theta$
for the example orbit.} 
\end{figure*}

In our 2-DOF case, the Near Identity Transformation [Eq.~(\ref{eq:nit})] is formulated as 
\be 
\begin{aligned}
\tilde q^\alpha &= q^\alpha+  \epsilon\sum_{n_r, n_\theta \in R} \frac{i}{n_r \Omega^r+n_\theta \Omega^\theta} \frac{\partial H_{n_r, n_\theta}}{\partial {J}_\alpha} e^{i (n_r q^r+n_\theta q^\theta)}\ , \\
\tilde{ {J}}_\alpha &=  {J}_\alpha +\epsilon \sum_{n_r, n_\theta\in R} \frac{n_\alpha}{n_r \Omega^r+n_\theta \Omega^\theta} H_{n_r, n_\theta}e^{i (n_r q^r+n_\theta q^\theta)}\ ,
\end{aligned}
\ee 
where $\{n_r, n_\theta\}$ is defined as the set of all non-resonant integers. 
The Near Identity Transformation involves operations on different Fourier components $H_{n_r, n_\theta}$  of the perturbed Hamiltonian $H_{\rm int}(q^\alpha, J_\beta)$ and its derivatives $\partial H_{n_r, n_\theta}/\partial J_\alpha$.

As a first step, we need to write the perturbed Hamiltonian in terms of action-angle variables 
\be 
H_{\rm int}(x^\mu, p_\nu)=\frac{1}{2} h^{\mu\nu}(x^A) p_\mu p_\nu \rightarrow H_{\rm int}(q^\alpha, J_\beta)\ ,
\ee 
i.e., we need to determine the inverse mapping $x^A(q^\alpha; J_\beta)$ and $p_A(q^\alpha; J_\beta)$, where $A\in\{r, \theta\}$. Making use of the symplectic relations between two sets of canonical variables
and the chain rule, we obtain 
\be \label{eq:symplectic}
\begin{aligned}
    \frac{\partial x^\mu}{\partial q^\alpha} &= \frac{\partial J_\alpha}{\partial p_\mu} = 
     \frac{\partial J_\alpha}{\partial P_\beta} \frac{\partial P_\beta}{\partial p_\mu}\ , \\
    \frac{\partial p_\mu}{\partial q^\alpha} &= -\frac{\partial J_\alpha}{\partial x^\mu}
    =-\frac{\partial J_\alpha}{\partial P_\beta}\frac{\partial P_\beta}{\partial x^\mu}\ .
\end{aligned}
\ee 
The above equations show that $r, \theta$ and $p_{r,\theta}$ explicitly depend on $q^{r,\theta}$ only, e.g.,
\be 
\begin{aligned}
\frac{\partial r}{\partial q^r} &= \frac{\partial J_r}{\partial P_\beta}\frac{\partial P_\beta}{\partial p_r}\ , \\
\frac{\partial r}{\partial q^\theta} &= \frac{\partial J_\theta}{\partial P_\beta}\frac{\partial P_\beta}{\partial p_r}    
\end{aligned}
\ee 
and similar for $\theta$ and $p_{r,\theta}$. Note that the derivative $\frac{\partial P_\beta}{\partial p_r}$ calculation 
is somewhat subtle, e.g., $\partial C/\partial p_r$ would be zero if we treat the factor $\mu^2$ in Eq.~(\ref{eq:Carter}) 
as a constant. In fact,  we should replace $\mu^2$ with $-2H(x,p)$, i.e., 
$\partial C/\partial p_r =a^2\cos^2\theta\times (-2\partial H(x,p)/\partial p_r) $.

Given initial conditions $(r, \theta, p_r, p_\theta)|_{(q^r, q^\theta)=(0,0)}$, one can evolve the above equations and
numerically obtain 
$(r, \theta, p_r, p_\theta)$ for any $(q^r, q^\theta)\in [0,2\pi]\times[0, 2\pi]$. As a check of the accuracy of the numerical algorithm used in solving Eqs.~(\ref{eq:symplectic}), 
one can use the fact that $(r, \theta, p_r, p_\theta)$ are periodic functions of $(q^r, q^\theta)$ with period $2\pi$. 
Consistent with the mapping from $(x, p)$ to $(q, J)$ explained in the previous section, 
where we set $q^r(r=r_-)=0$ and $q^\theta(\theta=\theta_-)=0$ ,
we choose the following initial conditions 
\be 
(r, \theta, p_r, p_\theta)|_{(q^r, q^\theta)=(0,0)} = (r_-, \theta_-, 0, 0)\ .
\ee
As an example, we consider a Kerr BH with spin $a=0.2$ and an orbit with $P_\alpha = (H, E, L, C) = (-0.5, 0.96, 3.5, 1.552)$,
i.e., eccentricity $e=0.364$, semi-latus rectum $p=9.821$, and $\cos^2(\theta_-)=0.112$. The inverse mapping
is shown in  Fig.~\ref{fig:app4}. With the inverse mapping, it is straightforward to numerically obtain the perturbed Hamiltonian $H_{\rm int}(q^r, q^\theta;J_\alpha)$, 
which can be decomposed as 
\be
{H}_{\rm int}(q^r, q^\theta;J_\alpha) = \sum_{n_r, n_\theta} H_{n_r, n_\theta}(J_\alpha) e^{i(n_r q^r+ n_\theta q^\theta)}\ .
\ee
Its derivatives can be numerically calculated as 
\be 
 \frac{\partial H_{n_r, n_\theta}}{\partial J_\alpha}=
 \frac{\partial H_{n_r, n_\theta}}{\partial P_\beta}\frac{\partial P_\beta}{\partial J_\alpha}\ ,
\ee 
when no analytic expression of $\frac{\partial H_{n_r, n_\theta}}{\partial P_\beta}$
is available.

\bibliography{ms}

\end{document}